\documentclass{article}

\usepackage[T1]{fontenc}
\usepackage{graphicx}

\usepackage{url}
\usepackage[hidelinks]{hyperref}
\usepackage[utf8]{inputenc}
\usepackage{booktabs}
\usepackage{algorithm}
\usepackage{algorithmic}
\usepackage{comment}

\usepackage{amsthm}
\usepackage{amssymb}
\usepackage{bbm}
\usepackage{amsmath}
\usepackage{amsfonts}
\usepackage{mathtools}
\usepackage{framed}
\usepackage{pifont}
\newcommand{\cmark}{\ding{51}}%
\newcommand{\xmark}{\ding{55}}%
\usepackage{xspace}

\usepackage{float}
\usepackage{tikz}
\usetikzlibrary{decorations.pathreplacing,patterns,positioning}

\usepackage{xcolor}

\usepackage{multirow}

\newcommand*{\task}[4]{
	\draw[rounded corners] (#3-#2,#4+0.1) rectangle (#3,#4+0.6);
	\draw (#3-#2/2,#4+0.35) node {#1};
}

\newtheorem{example}{Example}
\newtheorem{theorem}{Theorem}
\newtheorem{prop}{Proposition}
\newtheorem{definition}{Definition}
\newtheorem*{sketch}{Sketch of the proof}{\itshape}{\rmfamily}

\newtheorem*{sketchrq}{Sketch of the proof and remark}{\itshape}{\rmfamily}

\newcommand{\fixme}[1]{{\color{blue} [FixMe] {#1}}}
\newcommand{\sigmaD}{$\Sigma $D\xspace}
\newcommand{\sigmaT}{$\Sigma $T\xspace}
\newcommand{\sigmaDp}{{$\Sigma $$D_{P}$}\xspace}

\newtheorem{innercustomgeneric}{\customgenericname}
\providecommand{\customgenericname}{}
\newcommand{\newcustomtheorem}[2]{%
  \newenvironment{#1}[1]
  {%
   \renewcommand\customgenericname{#2}%
   \renewcommand\theinnercustomgeneric{##1}%
   \innercustomgeneric
  }
  {\endinnercustomgeneric}
}

\newcustomtheorem{customthm}{Theorem}
\newcustomtheorem{customprop}{Proposition}
\newcustomtheorem{customlemma}{Lemma}

\begin{document}
\title{Collective schedules: axioms and algorithms}
%
%
\author{Martin Durand \and
Fanny Pascual}
\date{}
%
%
%
\maketitle              
\begin{abstract}
The collective schedules problem consists in computing a schedule of tasks shared between individuals. Tasks may have different duration, and individuals have preferences over the order of the shared tasks. This problem has numerous applications since tasks may model public infrastructure projects, events taking place in a shared room, or work done by co-workers. Our aim is, given the preferred schedules of individuals (voters), to return a consensus schedule.  We propose an axiomatic study of the collective schedule problem, by using classic axioms in computational social choice and new axioms that take into account the duration of the tasks. We show that some axioms are incompatible, and we study the axioms fulfilled by three rules: one which has been studied in the seminal paper on collective schedules~\cite{pascual2018collective}, one which generalizes the Kemeny rule, and one which generalizes Spearman's footrule. From an algorithmic point of view, we show that these rules solve NP-hard problems, but that it is possible to solve optimally these problems for small but realistic size instances, and we give an efficient heuristic for large instances. We conclude this paper with experiments. 

\end{abstract}

\section{Introduction}
\label{sec:introduction}
In this paper, we are interested in the scheduling of tasks of interest to different people, who express their preferences regarding the order of execution of the tasks. The aim is to compute a consensus schedule which aggregates as much as possible the preferences of the individuals, that we will call voters in the sequel.  

This problem has numerous applications. For example, public infrastructure projects, such as extending the city subway system into several new metro lines, or simply rebuilding the sidewalks of a city, are often phased. Since  workforce, machines and yearly budgets are limited, phases have to be done one after the other. The situation is then as follows: given the different phases of the project (a phase being the construction of a new metro line, or of a new sidewalk), we have to decide in which order to do the phases. Note that phases may have different duration -- some may be very fast while some others may last much longer. In other words, the aim is to find a schedule of the phases, each one being considered as a task of a given duration. In order to get such a schedule, public authorities may take into account the preferences of citizens, or of citizens' representatives. 
Note that tasks may not only represent public infrastructure projects, but they may also model events taking place in a shared room, or work done by co-workers (the schedule to be built being the order in which the events -- or the work to be done -- must follow each other).

This problem, introduced in~\cite{pascual2018collective}, takes as input the preferred schedule of each voter (the order in which he or she would like the phases to be done), and returns one collective schedule -- taking into account the preferences of the voters and the duration of the tasks. 
We distinguish two settings. In the first one, each voter would like each task to be scheduled as soon as possible, even if he or she has preferences over the tasks. 
In other words, if this were possible, all the voters would agree to schedule all the tasks simultaneously as soon as possible. 
This assumption -- the earlier a task is scheduled the better -- ,  will be denoted  by \emph{EB} in the sequel. It was assumed  in~\cite{pascual2018collective}, and is reasonable in many situations, in particular when tasks are public infrastructure projects. However, it is not relevant in some other situations. 
Consider for example workers, or members of an association, who share different works that have to be done sequentially, for example because the tasks need the same workers, or the same resource (e.g. room, tool). Each work (task) has a given duration and can imply a different investment of each worker (investment or not of a person, professional travel, staggered working hours, ...). Each worker indicates his or her favorite schedule according to his or her personal constraints and preferences. In this setting, it is natural to try to fit as much as possible to the schedules wanted by the workers -- and scheduling a task much earlier than wanted by the voters is not a good thing. In this paper, our aim is to compute a socially desired collective schedule, with or without the EB assumption. 


This problem generalizes the consensus ranking problem, since if all the tasks have the same unit length, the preferred schedules of the voters can be viewed as preferred rankings of tasks. Indeed, each task can be considered as a candidate (or an item), and a schedule can be considered as a ranking of the candidates (items). Computing a collective schedule in this case consists thus in computing a collective ranking, a well-known problem in computational social choice. 


\medskip 

\noindent{\bf Related work.}
Our work is at the boundary between computational social choice~\cite{brandt2016handbook} and scheduling~\cite{brucker1999scheduling}, two major domains in artificial intelligence and operational research. 

As mentioned above, the collective schedule problem generalizes the collective ranking problem, which is an active field in computational social choice (see e.g. ~\cite{DworkKNS01,SkowronLBPE17,CelisSV18,Singh2018,Biega2018,Geyik2019,Asudeh2019,Narasimhan2020}). In this field, authors often design rules (i.e. algorithms) which return fair rankings, and they  often focus on fairness in the beginning of the rankings. If the items to be ranked are recommendations (or restaurants, web pages, etc.) for users, the beginning of the ranking is indeed probably the most important part. Note that this does not  hold for our problem since all the planned tasks will be executed -- only their order matters.  This means that rules designed for the collective ranking problem are not suitable not only because they do not consider duration for the items, but also because they focus on the beginning of the ranking. This also means that the rules we will study can be relevant for consensus ranking problems where the whole ranking is of interest.   


As mentioned earlier, the collective schedule problem has been introduced in~\cite{pascual2018collective} for the EB setting. In this paper, the authors introduced a weighted variant of the Condorcet principle, called the PTA Condorcet principle, where PTA stands for ``Processing Time Aware'' (cf. page~\pageref{def:pta_condorcet})
, and they adapted  previously known  Condorcet consistent rules when tasks have different processing times. They also introduced a new rule, which computes a schedule which minimizes the sum of the tardiness of tasks between the preferred schedules of the voters and the schedule which is returned. They show that the optimization problem solved by this rule  is NP-hard but that it can be solved for reasonable instances with a linear program.

Up to our knowledge, there is no other work which considers the schedule of common shared tasks between voters, or agents. Multi agent scheduling problems mainly focus on cases where (usually two) agents own their \emph{own} tasks, that are scheduled on shared machines: the aim is to find a Pareto-optimal and/or a fair schedule of the tasks of the agents, each agent being interested by her own tasks only~\cite{SauleT09,agnetis2014}. 

We conclude this related work section by mentioning similarities between our problem and the participatory budgeting problem, which is widely studied~\cite{Aziz2021}. In the participatory budgeting problem, voters give their preferences over a set of projects of different costs, and the aim is to select a socially desirable set of items of maximum cost $B$ (a given budget). 
The participatory  budgeting problem and the collective schedules problems  have common features. They both extend a classical optimization problem when users have preferences: the participatory  budgeting problem approach extends the knapsack problem when users have preferences over the items, while the collective schedules problem extends the scheduling problem when users have preferences on the order of the tasks.  Moreover, when considering unit items or tasks, both problems extend famous computational social choice problems: the participatory budgeting problem generalizes the multi winner voting problem when items have the same cost, and the collective schedules problem generalizes the collective ranking problem when tasks have the same duration. For both problems, because of the costs/lengths of the items/tasks, classical algorithms used with unit items/tasks may  return very bad solutions, and new algorithms are needed. 



\medskip 

\noindent{\bf Our contribution and map of the paper.}

\begin{itemize} 
\item In section~\ref{sec:preliminaries}, we present three rules to compute consensus schedules. We introduce the first one, that we will denote by PTA Kemeny, and which extends the well-known Kemeny rule used to compute consensus rankings in computational social choice~\cite{brandt2016handbook}. The two other rules come from scheduling theory, and were introduced in~\cite{pascual2018collective}: they consist in minimizing the sum of the tardiness of tasks in the returned schedule with respect to the voters' schedules  (rule \sigmaT), or in minimizing the sum of the deviation of tasks with respect to the voters' schedules (rule \sigmaD). Note that this last rule is equal to the Spearman's footrule~\cite{stanford1976spearman} when the tasks are unitary. 

\item In section~\ref{sec:axioms}, we study the axiomatic properties of the above mentioned rules by using classical social choice axioms as well as new axioms taking into account the duration of the tasks. 
Table~\ref{tab:axioms} summarizes our results. 
We also show incompatibilities between axioms: we show that a rule which is neutral, or which is based on a distance, both does not fulfill the PTA Condorcet consistency property, and can return a schedule with a sum of tardiness as far from the optimal as wanted. 

\item In Section~\ref{sec:resolution}, we show that the PTA Kemeny and \sigmaD rules solve NP-hard problems and we propose a fast heuristic which approximates the \sigmaD rule.

\item In Section~\ref{sec:experiments}, we see that the PTA Kemeny and \sigmaD rules can be used for small but realistic size instances, and that the heuristic presented in the previous section returns  schedules which are very close to the ones returned by \sigmaD. We also compare the performance of the three rules on the sum of tardiness or deviations of the tasks in the returned schedules. 
\end{itemize}
\vspace{-0.5em}
Let us now introduce formally our problem and present the three rules that we will study in the sequel.

%
%
%
%


\vspace{-0.5em}
\section{Preliminaries}
\label{sec:preliminaries}

\noindent{\bf Definition of the problem and notations.} 
Let $\mathcal{J}\!=\!\{1, \dots, n\}$ be a set of $n$ tasks. Each task $i\!\in\!\mathcal{J}$ has a length (or processing time) $p_i$. 
We do not consider idle times between the tasks, and preemption is not allowed: a schedule of the tasks is thus a permutation of the tasks of $\mathcal{J}$. We denote by $X^{\mathcal{J}}$ the set of all possible schedules. 
We denote by $V\!=\!\{1, \dots, v\}$ the set of $v$ voters. Each voter $k\!\in\! V$ expresses her favorite schedule $\mathcal{V}_k \!\in\! X^{\mathcal{J}}$ of the tasks in $\mathcal{J}$. The preference profile, $P$, is the set of these schedules: $P=\{\mathcal{V}_1, \dots, \mathcal{V}_v\}$. 

Given a schedule $S$, we denote by $C_i(S)$ the completion time of task $i$ in $S$. We denote by $d_{i,k}$ the completion time of task $i$ in the preferred schedule of voter $k$ (i.e. $d_{i,k}\!=\!C_i(\mathcal{V}_k)$) -- here $d$ stands for ``due date'' as this completion time can be seen as a due date, as we will see in the sequel. We denote by $a\succ_{S}b$ the fact that task $a$ is scheduled before task $b$ in schedule $S$. This relation is transitive, therefore, if, in a schedule $S$, task $a$ is scheduled first, then task $b$ and finally task $c$, we can describe $S$ as ($a \succ_S b \succ_S c)$. 

An \emph{aggregation rule} is a mapping $r\!:\!(X^{\mathcal{J}})^v\!\rightarrow\!X^{\mathcal{J}}$ that associates a schedule $S$ -- the consensus schedule -- to any preference profile $P$. We will focus on three aggregation rules that we introduce now: \emph{\sigmaD}, \emph{\sigmaT} and PTA Kemeny. 

\medskip

\noindent{\bf Three aggregation rules.} 

\medskip

\noindent{\bf A) The \sigmaD rule.} 
The \sigmaD rule 
is an extension of the Absolute Deviation (D) scheduling metric \cite{brucker1999scheduling}. This metric measures the deviation between a schedule $S$ and a set of given due dates for the tasks of the schedule. It sums, over all the tasks, the absolute value of the difference between the completion time of a task $i$ in $S$ and its due date. 
By considering the completion time $d_{i,k}$ of task $i$ in the preferred schedule $\mathcal{V}_k$ as a due date given by voter $k$ for task $i$, we express the deviation $D(S, \mathcal{V}_k)$ between  schedule $S$ and schedule $\mathcal{V}_k$ as $D(S,\mathcal{V}_k)=\sum_{i \in \mathcal{J}} |C_i(S)-d_{i,k}|$.
By summing over all the voters, we obtain a metric $D(S, P)$ measuring the deviation between a schedule $S$ and a preference profile $P$:
\vspace{-0.7em}
\begin{equation}
    D(S,P)=\sum\limits_{\mathcal{V}_k \in P} \sum\limits_{i \in \mathcal{J}} |C_i(S)-d_{i,k}|
\label{eq:formule_sigmaD}
\end{equation}

The \sigmaD rule returns a schedule $S^*$ minimizing the deviation with the preference profile $P$: $D(S^*,P)\!=\!\min_{S \in X^\mathcal{J}}D(S, P)$. 

This rule was introduced (but not studied) in \cite{pascual2018collective}, where the authors observed that, if tasks have unitary lengths, this rule minimizes the Spearman distance, which is defined as $S(S,\mathcal{V}_k)\!=\!\sum_{i \in \mathcal{J}} |\mathbf{pos}_i(S)-\mathbf{pos}_i(\mathcal{V}_k)|$, where $\mathbf{pos}_j(S)$ is the position of item $j$ in ranking $S$, i.e. the completion time of task $j$ in schedule $S$ if items are unitary  tasks.

\medskip

\noindent{\bf B) The \sigmaT rule.} 
This rule, introduced in \cite{pascual2018collective}, extends the classical Tardiness (T) scheduling criterion \cite{brucker1999scheduling}. The tardiness of a task $i$ in a schedule $S$ is 0 if task $i$ is scheduled in $S$ before its due date, and is equal to its delay with respect to its due date otherwise. As done for \sigmaD, we consider the completion time of a task $i$ in schedule $\mathcal{V}_k$ as the due date of voter $k$ for task $i$. The sum of the tardiness of the tasks in a schedule $S$ compared to the completion times in a preference profile $P$ is then:
\vspace{-1.35em}
\begin{equation}
    T(S,P)=\sum\limits_{\mathcal{V}_k \in P} \sum\limits_{i \in \mathcal{J}} \max(0,C_i(S)-d_{i,k})
    \label{eq:formule_sigmaT}
\end{equation}
\vspace{-1.1em}

The \sigmaT rule returns a schedule minimizing the sum of tardiness with $P$. 
\medskip

\noindent{\bf C) The PTA Kemeny rule.} 
We introduce a new rule, the Processing Time Aware Kemeny rule, an extension of the well-known \emph{Kemeny rule}~\cite{kemeny1959mathematics}. The Kendall tau distance is a famous metric to measure how close two rankings are: it counts the number of pairwise disagreements between two rankings (for each pair of candidates  $\{a,b\}$ it counts one if $a$ is ranked before $b$ in one ranking and not in the other ranking). 
The Kemeny rule minimizes the sum of the Kendall tau distances  to the preference profile, i.e. the voter's preferred rankings.  

Despite its good axiomatic properties, this rule, which does not take into account the length of the tasks, is not suitable for our collective schedules problem.  Consider for example an instance with only two tasks, a short task $a$ and a long task $b$. If a majority of voters prefer $b$ to be scheduled first, then in the returned schedule it will be the case. However, in EB settings, it may be suitable that $a$ is scheduled before $b$ since the small task $a$ will delay the large one $b$ only by a small amount of time, while the contrary is not true.

We therefore propose a weighted extension of the Kemeny rule: the PTA Kemeny rule, which minimizes the sum of weighted Kendall tau distances between a schedule $S$ and the schedules of the preference profile $P$. The weighted Kendall tau distance between two schedules $S$ and $\mathcal{V}_k$  counts  the weighted number of pairwise disagreements between two rankings; for each pair of tasks  $\{a,b\}$ such that $b$ is scheduled before $a$ in  $\mathcal{V}_k$ and not in $S$, it counts $p_a$. 
This weight measures the delay caused by task $a$ on task $b$ in $S$ (whereas $a$ caused no delay on $b$ in $\mathcal{V}_k$). 
The score measuring the difference between a schedule $S$ and $P$ is:




\vspace{-0.8em}
\begin{equation}
    \Delta_{PTA}(S,P)=\sum\limits_{\mathcal{V}_k \in P} \sum\limits_{\{a,b\} \in C^2} \mathbbm{1}_{a \succ_S b, b \succ_{\mathcal{V}_k} a}\times p_a
    \label{eq:formule_pta_kemeny}
\end{equation}
\vspace{-0.8em}




\begin{example} 
We consider an instance with three tasks $\{1,2,3\}$ and five voters. We have $p_1=2, p_2=4$ and $p_3=1$. The preference profile is as follows (we indicate in front of each schedule the number of voters for which it is the preferred schedule):
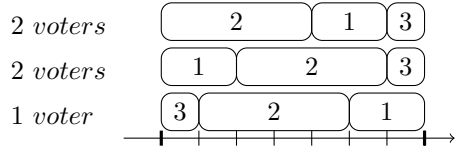
\begin{figure}[H]
\centering
\begin{tikzpicture}
\task{$2$}{2}{2}{1.2}
\task{$1$}{1}{3}{1.2}
\task{$3$}{0.5}{3.5}{1.2}

\task{$1$}{1}{1}{0.6}
\task{$2$}{2}{3}{0.6}
\task{$3$}{0.5}{3.5}{0.6}

\task{$3$}{0.5}{0.5}{0}
\task{$2$}{2}{2.5}{0}
\task{$1$}{1}{3.5}{0}

\draw[->](0-0.5,0)--(3.5+0.4,0);

\node[text width=1cm] at (-1.5,1.5) {$2\text{ }voters$};
\node[text width=1cm] at (-1.5,0.9) {$2\text{ }voters$};
\node[text width=1cm] at (-1.5,0.3) {$1\text{ }voter$};

\foreach \x in {0.5,1,1.5,2,2.5,3,3.5}  
	\draw (\x,0.1)--(\x,-0.1);

\draw[line width=1.2pt] (0,0.1)--(0,-0.1);
\draw[line width=1.2pt] (3.5,0.1)--(3.5,-0.1);

\end{tikzpicture}
\label{fig:example_pta_kemeny}
\end{figure}

Let us compute the PTA Kendall tau score of schedule $S$, in which $2\!\succ\!1\!\succ\!3$ (where $a\!\succ\!b$ means that $a$ is scheduled before $b$).
There is 0 disagreement with the first set of 2 voters.
There is 1 disagreement with the second set of 2 voters because the pair $\{1,2\}$ is inverted. Therefore we count $p_2\!\times\!2\!=\!8$, since $2$ is ranked higher in $S$ than $1$. 
There are 2 disagreements with the last voter, one on the pair $\{1,3\}$ and one on the pair $\{2,3\}$. Therefore, we count $p_1\!=\!2$ plus $p_2\!=\!4$. 
Overall, the score of $S$ : $2\!\succ\!1\!\succ\!3$ is $8\!+\!2\!+\!4\!=\!14$.
\label{ex:PTA Kemeny}
\end{example}

\noindent{\bf Resoluteness.} Note that each of these rules returns a schedule minimizing an optimization function, and that it is possible that several optimal schedules exist. In computational social choice, rules may be partitioned into two sets: resolute and irresolute rules.  A rule is \emph{resolute} if it always returns one single solution, and it is \emph{irresolute} if it returns a set of solutions. Thus, rules optimizing an objective function may either be irresolute, and return all the optimal solutions, or they can be resolute and use a tie-breaking mechanism which allows to determine a unique optimal solution for each  instance. 

Irresolute rules have the advantage that a decision maker can choose among the optimal solutions, the one that he or she prefers. However, the set of optimal solutions can be large, and sometimes even exponential, making it difficult to compute in practice. Furthermore, in real situations, there is not always a decision maker which makes choices, and an algorithm has to return a unique solution: in this case, the rule must be resolute and needs to use a tie breaking rule that allows to decide between the optimal solutions. 

In this paper, we  consider that each rule returns a unique solution. However, since a good tie breaking mechanism is usually dependent on the context, we will not describe it. Instead, we will study the properties of the set of optimal solutions and see if using a tie breaking mechanism impacts the axiomatic properties of the rule -- as we will see, most of the time, this will not be the case. 

\vspace{-0.5em}
\section{Axiomatic properties}
\label{sec:axioms}



\vspace{-0.5em}
\subsection{Neutrality and PTA neutrality.}

The neutrality axiom is a classical requirement of a social choice rule. A rule is \emph{neutral} if it does not discriminate apriori between different candidates. Note that this axiom can be fulfilled only by irresolute rules, since a resolute rule should return only one solution, even when there are only two equal length tasks  $a$ and $b$, and two voters: one voter who prefers that $a$ is before $b$, while the other voter prefers that $b$ is before $a$ (the same remark holds for consensus rankings instead of consensus schedules).  Therefore, in this subsection we will consider that our three rules return all the optimal solutions of the function they optimize. 

\begin{definition}\textbf{\emph{(Neutrality)}} Let $r$ be an aggregation rule, $P$ a preference profile, and $\mathcal{S}^*$ the set of solutions returned by an irresolute rule $r$ when applied on $P$. Let $P_{(a \leftrightarrow b)}$ be the preference profile obtained from $P$ by switching the positions of two candidates (tasks) $a$ and $b$ in all the preferences and $\mathcal{S}^*_{(a \leftrightarrow b)}$ the set of solutions returned by $r$ on $P_{(a \leftrightarrow b)}$. The rule $r$ is \emph{neutral} iff, for each solution $S$ in $\mathcal{S}^*$, there exists a solution $S_{(a \leftrightarrow b)}$ in $\mathcal{S}^*_{(a \leftrightarrow b)}$, such that $S_{(a \leftrightarrow b)}$ can be obtained from $S$ by swapping the positions of $a$ and $b$. 
\end{definition}


\begin{prop}
The \sigmaD rule is not neutral even if it does not apply any tie-breaking mechanism.
\end{prop}

\begin{proof}
Let us consider an instance with $n=6$ tasks $\{a,b,c,d,e,f\}$, we have $p_a=p_b=p_c=1,p_d=2,p_e=k$ and $p_f=k-2$, with $k$ a positive integer. The instance has a high even number of $v$ voters having the following preferences:

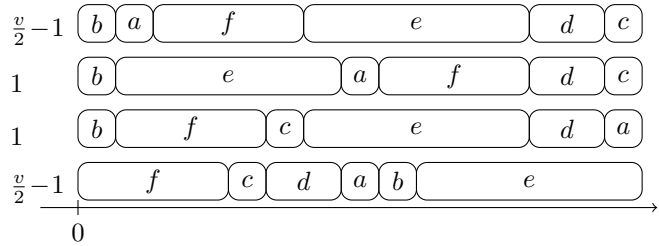
\begin{figure}[H]
\centering
\begin{tikzpicture}
\task{$b$}{0.5}{0.5}{2.1}
\task{$a$}{0.5}{1}{2.1}
\task{$f$}{2}{3}{2.1}
\task{$e$}{3}{6}{2.1}
\task{$d$}{1}{7}{2.1}
\task{$c$}{0.5}{7.5}{2.1}

\task{$b$}{0.5}{0.5}{1.4}
\task{$e$}{3}{3.5}{1.4}
\task{$a$}{0.5}{4}{1.4}
\task{$f$}{2}{6}{1.4}
\task{$d$}{1}{7}{1.4}
\task{$c$}{0.5}{7.5}{1.4}

\task{$b$}{0.5}{0.5}{0.7}
\task{$f$}{2}{2.5}{0.7}
\task{$c$}{0.5}{3}{0.7}
\task{$e$}{3}{6}{0.7}
\task{$d$}{1}{7}{0.7}
\task{$a$}{0.5}{7.5}{0.7}

\task{$f$}{2}{2}{0}
\task{$c$}{0.5}{2.5}{0}
\task{$d$}{1}{3.5}{0}
\task{$a$}{0.5}{4}{0}
\task{$b$}{0.5}{4.5}{0}
\task{$e$}{3}{7.5}{0}

\draw[->](0-0.5,0)--(7.5+0.2,0);
\draw (0, 0.1)--(0,-0.1) node[below]{$0$};

\node[text width=1cm] at (-0.4,2.35) {$\frac{v}{2}\!-\!1$};
\node[text width=1cm] at (-0.4,1.65) {$1$};
\node[text width=1cm] at (-0.4,0.95) {$1$};
\node[text width=1cm] at (-0.4,0.25) {$\frac{v}{2}\!-\!1$};
\end{tikzpicture}
\label{fig:profile_couter_example_neutrality}
\end{figure}

For $k=20$ and $v=400$, the \sigmaD rule returns the schedule $S$ ($b \succ f \succ a \succ e \succ d \succ c$) since it is the only one minimizing the absolute deviation with the profile. If we consider the same profile but in which the positions of $b$ and $e$ are swapped, we have:

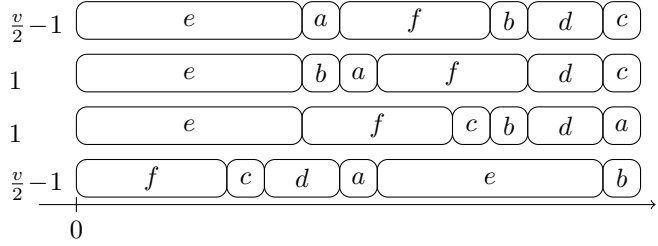
\begin{figure}[H]
\centering
\begin{tikzpicture}
\task{$e$}{3}{3}{2.1}
\task{$a$}{0.5}{3.5}{2.1}
\task{$f$}{2}{5.5}{2.1}
\task{$b$}{0.5}{6}{2.1}
\task{$d$}{1}{7}{2.1}
\task{$c$}{0.5}{7.5}{2.1}

\task{$e$}{3}{3}{1.4}
\task{$b$}{0.5}{3.5}{1.4}
\task{$a$}{0.5}{4}{1.4}
\task{$f$}{2}{6}{1.4}
\task{$d$}{1}{7}{1.4}
\task{$c$}{0.5}{7.5}{1.4}

\task{$e$}{3}{3}{0.7}
\task{$f$}{2}{5}{0.7}
\task{$c$}{0.5}{5.5}{0.7}
\task{$b$}{0.5}{6}{0.7}
\task{$d$}{1}{7}{0.7}
\task{$a$}{0.5}{7.5}{0.7}

\task{$f$}{2}{2}{0}
\task{$c$}{0.5}{2.5}{0}
\task{$d$}{1}{3.5}{0}
\task{$a$}{0.5}{4}{0}
\task{$e$}{3}{7}{0}
\task{$b$}{0.5}{7.5}{0}

\draw[->](0-0.5,0)--(7.5+0.2,0);
\draw (0, 0.1)--(0,-0.1) node[below]{$0$};

\node[text width=1cm] at (-0.4,2.35) {$\frac{v}{2}\!-\!1$};
\node[text width=1cm] at (-0.4,1.65) {$1$};
\node[text width=1cm] at (-0.4,0.95) {$1$};
\node[text width=1cm] at (-0.4,0.25) {$\frac{v}{2}\!-\!1$};
\end{tikzpicture}
\end{figure}
For profile $P'$, the only optimal schedule $S'$ is ($e \succ a \succ f \succ b \succ d \succ c$). If the \sigmaD rule was neutral, $S$ and $S'$ would be similar but the position of $b$ and $e$ would be swapped. However, the positions of $a$ and $f$ are also modified, meaning that the \sigmaD rule is not neutral.
\end{proof}

As we will see later, the \sigmaT and the PTA Kemeny rules do not fulfill neutrality (this will be corollaries of Propositions \ref{prop:neutrality_pta} and \ref{prop:neutrality_sigmaT}).

Since neutrality leads to unsatisfactory solutions, and since we want an equal treatment between comparable tasks, we introduce the \emph{PTA neutrality} axiom, which ensures that two tasks of equal length are considered in the same way. 


\begin{definition}\textbf{\emph{(PTA neutrality)}}
Let $r$ be an aggregation rule, $P$ a preference profile, and $\mathcal{S}^*$ the set of solutions returned by an irresolute rule $r$ when applied on $P$. Let $P_{(a \leftrightarrow b)}$ be the preference profile obtained from $P$ by switching the positions of two tasks $a$ and $b$ in all the preferences and $\mathcal{S}^*_{(a \leftrightarrow b)}$ the set of solutions returned by $r$ on $P_{(a \leftrightarrow b)}$. The rule $r$ is \emph{PTA neutral} iff, for any two tasks $a$ and $b$ such that $p_a=p_b$, for each solution $S$ in $\mathcal{S}^*$, there exists a solution $S_{(a \leftrightarrow b)}$ in $\mathcal{S}^*_{(a \leftrightarrow b)}$, such that $S_{(a \leftrightarrow b)}$ can be obtained from $S$ by swapping the positions of $a$ and $b$. 
\end{definition}

The PTA neutrality axiom extends the concept of neutrality for the cases in which tasks (candidates) have lengths (weights). This axiom ensures that two candidates with the same characteristics are treated equally. 
When all the tasks have the same length, the PTA neutrality axiom is equal to the neutrality axiom. 

\begin{prop}
The PTA Kemeny, \sigmaD and \sigmaT rules are PTA neutral if they do not apply any tie-breaking mechanism.
\end{prop}

\begin{proof}
The PTA Kemeny, \sigmaD and \sigmaT rules all return schedules that minimize an objective function. For \sigmaD and \sigmaT, this objective function depends on the completion times of all tasks in both the returned schedule and in the preferences in the profile. Let us swap two tasks $i$ and $j$ with the same length in the preferences of the profile $P=\{\mathcal{V}_1, \dots, \mathcal{V}_v\}$, giving us a profile $P'=\{\mathcal{V}'_1, ..., \mathcal{V}'_v\}$. We have $C_i(\mathcal{V}'_k)=C_j(\mathcal{V}_k)$ and $C_j(\mathcal{V}'_k)=C_i(\mathcal{V}_k)$ since no other task has been moved and since $i$ and $j$ have the same length. For each possible consensus schedule $S$ for $P$, we can note that the schedule similar to $S$ but in which $i$ and $j$ are swapped has the same deviation (resp. tardiness) for $P'$ than $S$ for $P$. This implies that if a schedule $S$ is optimal for a profile $P$, then the schedule $S'$ obtained by swapping $i$ and $j$ is optimal for the profile $P'$: the PTA neutrality holds for \sigmaD and \sigmaT. 

The PTA Kemeny rule returns a ranking minimizing the weighted sum of pairwise disagreements with the profile. By swapping the positions of $i$ and $j$ in both the profile $P$ and a schedule $S$, we do not change the disagreements on pairs of tasks that do not contain $i$ or $j$. The pair $\{i,j\}$ is permuted in both $S$ and $P$ leading to the same number of disagreements. Whenever there was a disagreement between $S$ and a preference in $P$ on a pair $\{i,x\}$ then there will be a disagreement on $\{j,x\}$ between $S'$ and $P'$. Similarly if $S$ and a preference of $P$ agree on the pair $\{i,x\}$, then $S'$ and the corresponding preference in $P'$ will agree on $\{j,x\}$. Since the lengths of $i$ and $j$ are the same, the weights on a disagreement between $S$ and $P$ will be the same than a disagreement between $S'$ and $P'$, meaning that the overall sum of weighted disagreements between $S$ and $P$ is the same than between $S'$ and $P'$. Since this applies to every schedule $S$, if $S$ is optimal for the profile $P$, $S'$ is optimal for $P'$, hence the PTA Kemeny rule is PTA neutral.
\end{proof}

Note that the neutrality axiom is incompatible with the resoluteness axiom \cite{brandt2016handbook}. That means that any rule returning always only one solution cannot be neutral. For our rules, once we use a tie-breaking mechanism, we have to give up neutrality. However, if we focus on the set of optimal solutions, it could respect neutrality.

\subsection{Distance.}
Some aggregation rules are based on the minimization of a metric. By metric, we mean a mapping between a pair of elements, most of the time a preference and a solution, and a value. Most of these rules then sum these values over the whole preference profile to evaluate the difference between a solution and a preference profile. For example, the \sigmaT rule returns a schedule minimizing the sum of tardiness with the preferences of the profile. If the metric is a distance (i.e. it satisfies non-negativity, identity of indiscernible, triangle inequality and symmetry), we say that the aggregation rule is ``based on a distance''.

\begin{prop}
The absolute deviation metric is a distance.
\label{prop:sigmad_distance}
\end{prop}

\begin{proof}
To be a distance, a metric $m$ must fulfill four properties:
\begin{itemize}
	\item[(1)] Non negativity: $m(S,S') \geq 0, \forall S,S' \in X^{\mathcal{J}}$
	\item[(2)] Identity of indiscernibles: $m(S,S')=0$ iff $S=S'$
	\item[(3)] Symmetry: $m(S,S')=m(S',S), \forall S,S' \in X^{\mathcal{J}}$
	\item[(4)] Triangle inequality: $m(S,S') \leq m(S,z) +m(z,S'), \forall S,S',z \in X^{\mathcal{J}}$
\end{itemize}
The non negativity (1) property is direct since we sum absolute values, which are always positive.\\
We prove the identity of indiscernibles (2) by noting that two schedules $S,S'$ are identical iff all the tasks complete at the exact same time in both schedules. Therefore, if $S$ and $S'$ are identical, then there is no difference between the completion times of a task in the two schedules, and the deviation is thus null. Otherwise, at least one task completes at a different time in the two schedules, leading to a non-null difference, and a positive overall absolute deviation between the two schedules.\\ 
The symmetry (3) property is a direct consequence of the evenness of the absolute value function. 
By definition, $D(S,S')=\sum_{i \in \mathcal{J}} |C_i(S)-C_i(S')|$ and  $D(S',S)=\sum_{i \in \mathcal{J}} |C_i(S')-C_i(S)|$. By noting that: $C_i(S)-C_i(S')=-(C_i(S')-C_i(S))$ and since $|a|=|-a|, \forall a \in \mathbb{R}$, we have $D(S,S')=D(S',S)$.\\
Finally, we prove the triangle inequality (4) thanks to the subadditivity property of the absolute value function. We consider the absolute deviation between two schedules $S$ and $S'$: $D(S,S')$. Let $z$ be a third schedule. By definition:
$D(S,S')=\sum_{i \in \mathcal{J}} |C_i(S)-C_i(S')|=\sum_{i \in \mathcal{J}} |C_i(S)-C_i(z)+C_i(z)-C_i(S')|$. By subadditivity of the absolute value, we have:
\[
D(S,S') \leq \sum_{i \in \mathcal{J}} \left( |C_i(S)-C_i(z)|+|C_i(z)-C_i(S')| \right)
\]
\[
D(S,S') \leq D(S,z)+D(z,S')
\]
\end{proof}

As we will see in the sequel (Propositions~\ref{prop:distance_pta} and~\ref{prop:distance_sigmaT}), the fact that the $D$ metric is a distance implies  that the \sigmaD rule is not PTA Condorcet consistent, and that it can return  solutions with a sum of tardiness arbitrarily larger than the optimal sum of tardiness. 
Before seeing this, let us start by recalling what is the PTA Condorcet consistency property, introduced in~\cite{pascual2018collective}.

\subsection{PTA Condorcet consistency.}


\begin{definition}[PTA Condorcet consistency  \cite{pascual2018collective}]
A schedule $S$ is PTA Condorcet consistent with a preference profile $P$ if, for any two tasks $a$ and $b$, it holds that $a$ is scheduled before $b$ in $S$ whenever at least $\frac{p_a}{p_a+p_b} \cdot v$ voters put $a$ before $b$ in their preferred schedule. A scheduling rule satisfies the PTA Condorcet principle if for each preference profile it returns only the PTA Condorcet consistent schedule, whenever such a schedule exist.
\label{def:pta_condorcet}
\end{definition}

Note that if all the tasks have the same length, the PTA Condorcet consistency is equal to the well-known Condorcet consistency~\cite{condorcet1785essai}.

\begin{prop}
The PTA Kemeny rule is PTA Condorcet consistent. 
\label{prop:PTA_Kemeny_Condorcet_consistency}
\end{prop}
\begin{proof}
Let  $S$ be a schedule returned by the PTA Kemeny rule. For the sake of contradiction, let us suppose that, in $S$, there is a pair of tasks $a$ and $b$ such that $a$ is scheduled before $b$ whereas more than $\frac{p_b}{p_a+p_b} \cdot v$ voters scheduled $b$ before $a$ and that a PTA Condorcet schedule exists.\\
We study two cases. Firstly, consider the tasks $a$ and $b$ are scheduled consecutively in $S$. In that case, we call $S_{(a \leftrightarrow b)}$ the schedule obtained from $S$ in which we swap the position of $a$ and $b$. Since both schedules are identical except for the inversion of the pair $\{a,b\}$ their weighted Kendall tau scores vary only by the number of disagreements on this pair.
\begin{itemize}
	\item  We have assumed that the number $v_b$ of voters scheduling $b$ before $a$ is larger than $\frac{p_b}{p_a+p_b} \cdot v$. Since in $S$, $a$ is scheduled before $b$, the weighted disagreement of the voters  on pair $\{a,b\}$ in $S$ is larger than $\frac{p_b}{p_a+p_b} \cdot v \cdot p_a$.
	\item In $S_{(a \leftrightarrow b)}$, $b$ is scheduled before $a$. 
	Since $v_b>\frac{p_b}{p_a+p_b} \cdot v$, we know that the number $v_a$ of voters scheduling $a$ before $b$ is smaller than $\frac{p_a}{p_a+p_b} \cdot v$. Therefore, the weighted disagreement on pair $\{a,b\}$ is smaller than $\frac{p_a}{p_a+p_b} \cdot v \cdot p_b$.
\end{itemize}
Thus the score of $S_{(a \leftrightarrow b)}$ is smaller than the one of $S$: $S$ is not optimal for the PTA Kemeny rule, a contradiction.

Secondly, let us consider that $a$ and $b$ are not consecutive in $S$, and let $c$ be the task which follows $a$ in $S$. 
In $S$, it is not possible to swap two consecutive tasks to reduce the weighted Kendall tau score, otherwise the schedule could not be returned by the PTA Kemeny rule. 
Thus, by denoting by $S_{(a \leftrightarrow c)}$ the schedule $S$ in which we exchange the order of tasks $a$ and $c$, we get that $\Delta_{PTA}(S_{(a \leftrightarrow c)},P)-\Delta_{PTA}(S,P)\geq 0$. This implies that $v_a \cdot p_c-v_c \cdot p_a\geq 0$ and $v_a \cdot \frac{p_c}{p_a+p_c}-v_c \cdot \frac{p_a}{p_a+p_c}\geq 0$, 
where $v_c=v-v_a$ is the number of voters who schedule $c$ before $a$ in their preferred schedule. Therefore, $v_a\geq v \cdot \frac{p_a}{p_a+p_c}$. Therefore, 
task $a$ is scheduled before $c$ in any PTA Condorcet consistent schedule. By using the same argument, we find that task $c$ is scheduled before task $d$, which follows $c$ in $S$, and that $c$ has to be scheduled before $d$ in any PTA Condorcet consistent schedule, 
and so forth until we meet task $b$. 
This set of tasks forms a cycle since $a$ has to be scheduled before $c$ in a PTA Condorcet consistent schedule, $c$ has to be scheduled before $d$ in a PTA Condorcet consistent schedule, $\dots$, until we met $b$. Moreover $b$ has to be scheduled before $a$ in a PTA Condorcet consistent schedule since more than $\frac{p_b}{p_a+p_b} \cdot v$ voters scheduled $b$ before $a$. The existence of this cycle means that no PTA Condorcet consistent schedule exists for the profile, a contradiction.
\end{proof}

\subsection{Incompatibilities between axioms and properties.}

One can wonder if the PTA Kemeny rule (without breaking-tie rule) is the only rule which is PTA Condorcet consistent, neutral and which fulfills  reinforcement, just like the Kemeny rule is the only Condorcet consistent neutral rule fulfilling reinforcement \cite{young1978consistent}. We will show that it is not true, since PTA Kemeny does not fulfill neutrality. 
We even show a more general statement : no neutral rule can be PTA Condorcet consistent. This answers an open question of~\cite{pascual2018collective} where the author conjectured ``that rules satisfying neutrality and reinforcement fail the PTA Condorcet principle'' and said that ``it is an interesting open question whether such an impossibility theorem holds''.

\begin{prop}
No neutral rule can be PTA Condorcet consistent. 
\label{prop:neutrality_pta}
\end{prop} 

\begin{proof}
Let us consider an instance $I$ with an odd number of voters $v\geq 3$, two tasks $a$ and $b$, such that $p_a=1$ and $p_b=v$, and a preference profile as follows: 
$v_a=\frac{v-1}{2}$ voters prefer schedule $a \succ b$ (this schedule will be denoted by $A$), and $v_b=\frac{v+1}{2}$ voters prefer schedule $b \succ a$ (schedule denoted by $B$). 

By contradiction, let us suppose that  $r$ is a rule which is both neutral and PTA Condorcet consistent. 
Since $r$ is PTA Condorcet consistent, it will necessarily return the only PTA Condorcet consistent schedule when applied on instance $I$: $A$ (indeed at least $\frac{p_a}{p_a+p_b} \cdot v=\frac{v}{v+1}\leq 1$ voter prefer to schedule $a$ before $b$). 

Let $P_{(a \leftrightarrow b)}$ be the preference profile obtained from $P$ in which the positions of $a$ and $b$ are swapped in all the voters' preferences. Since $r$ is neutral, it  necessarily returns schedule $A$ in which we have inverted $a$ and $b$, i.e. schedule $B$. However, this schedule is not PTA Condorcet consistent, whereas there exists a PTA Condorcet schedule. Indeed, schedule $A$ is  a  PTA Condorcet consistent schedule for $P_{(a \leftrightarrow b)}$ since at least $\frac{p_a}{p_a+p_b} \cdot v=\frac{v}{v+1}\leq 1\leq v_a$ voters prefer to schedule $a$ before $b$, while $\frac{p_b}{p_a+p_b} \cdot v=\frac{v^2}{(v+1)}$ is larger than $v_b$ for all values of $v\geq 3$.  
 \qed
\end{proof}



This proposition implies that the PTA Kemeny rule is not neutral, even if no tie-breaking mechanism is used, since it is PTA Condorcet consistent.

Aggregation rules based on distance metrics have several good axiomatic properties \cite{brandt2016handbook}. However, we show that they cannot be PTA Condorcet consistent. 
Propositions~\ref{prop:distance_pta},~\ref{prop:neutrality_sigmaT},  and~\ref{prop:distance_sigmaT} can be proven in an analogous way to Proposition~\ref{prop:neutrality_pta}.

\begin{prop}
Any resolute aggregation rule returning a schedule minimizing a distance with the preference profile violates the PTA Condorcet consistency property. This result holds for any tie-breaking mechanism.
\label{prop:distance_pta}
\end{prop}

\begin{proof}
A distance relation $d$ respects symmetry, therefore, for each pair of schedules $S$ and $S'$, $d(S,S')=d(S',S)$.

Let us consider an instance $I$ with two tasks $a$ and $b$, such that $p_a=1$ and $p_b=v$, an odd number of voters $v\geq 3$, and a preference profile as follows: 
$v_a=\lfloor\frac{v-1}{2}\rfloor$ voters prefer schedule $a \succ b$ (this schedule will be denoted by $A$), and $v_b=\lceil\frac{v+1}{2}\rceil$ voters prefer schedule $b \succ a$ (schedule denoted by $B$). 
By symmetry $d(A,B)=d(B,A)$. Since $v_b > v_a$, any aggregation rule $r$ based on minimizing a distance with the profile will return $B$ only. 
However, the only Condorcet consistent schedule is $A$. Since rule $r$ returns $B$, $r$ is not PTA Condorcet consistent.
\end{proof}

Let us now show that neutrality and distance minimization can lead to very inefficient solutions for tardiness minimization.

\begin{prop}
For any $\alpha \geq 1$, there is no neutral aggregation rule returning a set of solutions $\mathcal{S}$ such that all the solutions in $\mathcal{S}$ are $\alpha$-approximate for \sigmaT.
\label{prop:neutrality_sigmaT}
\end{prop}

\begin{proof}
Let us consider an instance $I_k$ with two tasks $a$ and $b$, such that $p_a=1$ and $p_b=k$, an odd number of voters $v\geq 3$, and a preference profile as follows: 
$v_a=\lfloor\frac{v-1}{2}\rfloor$ voters prefer schedule $a \succ b$ (this schedule will be denoted by $A$), and $v_b=\lceil\frac{v+1}{2}\rceil$ voters prefer schedule $b \succ a$ (schedule denoted by $B$). 
We define profile $P_{(a \leftrightarrow b)}$ as the profile $P$ in which tasks $a$ and $b$ are swapped in the preferences. Any neutral rule which returns $A$ (resp. $B$) in $P$ will return $B$ (resp. $A$) in $P_{(a \leftrightarrow b)}$. A neutral rule could also return $\{A,B\}$ 

For profile $P$, schedule $A$ has a sum of tardiness of $\lfloor \frac{v-1}{2} \rfloor$, since task $b$ is delayed by $1$ in comparison to schedule $B$. Schedule $B$ has a sum of tardiness of $\lceil \frac{v+1}{2} \rceil \times k$ since task $a$ is delayed by $k$ in comparison to schedule $A$.

Likewise, in profile $P_{(a \leftrightarrow b)}$, schedule $A$ has a sum of tardiness of $\lceil \frac{v+1}{2} \rceil$ , and schedule $B$ has a sum of tardiness of $\lfloor \frac{v-1}{2} \rfloor . k$. 

For both profiles $P$ and $P_{(a \leftrightarrow b)}$, schedule  $B$ has a total sum of tardiness $k$ times higher than the optimal (schedule $A$), which can be arbitrarily far from the optimal. Since a neutral rule $r$ returns $B$ either for profile $P$ or for profile $P_{(a \leftrightarrow b)}$, or both, and since $k$ can be as big as we want, the sum of tardiness of at least one schedule returned by $r$ can be arbitrarily far from the optimal.
\end{proof}

Since the \sigmaT rule, without tie-breaking mechanism, returns only optimal solutions for the tardiness minimization, this implies that the \sigmaT rule is not neutral. 

\begin{prop}
For any $\alpha \geq 1$, there is no aggregation rule based on a distance minimization and always returning at least one $\alpha$-approximate solution for \sigmaT.
\label{prop:distance_sigmaT}
\end{prop}

\begin{proof}
Let us consider an instance $I_k$ with two tasks $a$ and $b$, such that $p_a=1$ and $p_b=k$, an odd number of voters $v\geq 3$, and a preference profile as follows: 
$v_a=\lfloor\frac{v-1}{2}\rfloor$ voters prefer schedule $a \succ b$ (this schedule will be denoted by $A$), and $v_b=\lceil\frac{v+1}{2}\rceil$ voters prefer schedule $b \succ a$ (schedule denoted by $B$). 
Any distance $d$ is symmetric, therefore $d(A,B)=d(B,A)$. Any rule returning the schedule minimizing the distance with the profile will return $A$ (resp. $B$) if $A$ (resp. $B$) is more present than $B$ (resp. $A$) in the profile. Since a majority of voter prefer $B$, any rule based on a distance minimization returns $B$.
For profile  $P$, $A$ has a total sum of tardiness of $\lceil \frac{v+1}{2} \rceil  \times 1$ since task $b$ is delayed by $1$ in comparison to schedule $B$. On the other hand, $B$ has a total sum of tardiness of $\lfloor \frac{v-1}{2} \rfloor \times k$, since  task $a$ is delayed by $k$ in comparison to schedule $A$. 
Since $k$ can be as high as we want, the sum of tardiness in schedule $B$ can be arbitrarily far from the optimal sum of tardiness.
\end{proof}

\subsection{Length reduction monotonicity.}
Let us now introduce a new axiomatic property, which is close to the  {discount monotonicity} axiom~\cite{talmon2019framework} for the participatory budgeting problem. A rule $r$ satisfies the \emph{discount monotonicity} axiom if a project cannot be penalised because it is cheaper (i.e. if a project is selected by rule $r$ then it should also be selected by this rule if its price decreases, all else being equal). We propose a new axiom, that we call  \emph{length  reduction monotonicity}, and which states that the starting time of a task in a schedule cannot be delayed if its length decreases  (all else being equal). This axiom is particularly meaningful in EB settings, where all the voters would like all the tasks to be scheduled as soon as possible. 

\begin{definition}\textbf{\emph{(Length Reduction Monotonicity)}}
Let $S$ be the schedule returned by a  resolute rule $r$ on instance $I$. Assume that we decrease the length of a task $t$ in $I$, all else being equal. Let $S'$ be the schedule returned by $r$ on this new instance. Rule $r$ fulfills \emph{length reduction monotonicity} if task $t$ does not start later in $S'$ than in $S$.
\end{definition}


\begin{prop}
The \sigmaD rule does not fulfill length reduction monotonicity for any tie-breaking mechanism.
\label{prop:sigmad_monotonicity}
\end{prop}

\begin{proof}
Let us consider an instance with 5 tasks $\{1,2,3,x,p\}$ with $p_1=p_2=p_3=p_x=1$ and $p_p=10$. The preferences of the 400 voters are as follows:

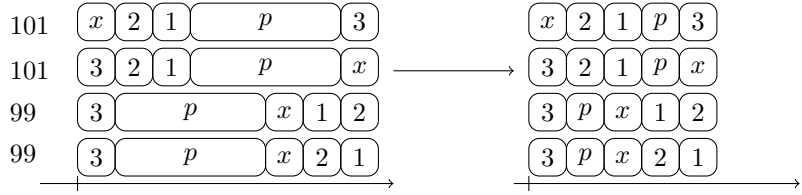
\begin{figure}[H]
\centering
\begin{tikzpicture}
\task{$x$}{0.5}{0.5}{1.8}
\task{$2$}{0.5}{1}{1.8}
\task{$1$}{0.5}{1.5}{1.8}
\task{$p$}{2}{3.5}{1.8}
\task{$3$}{0.5}{4}{1.8}

\task{$3$}{0.5}{0.5}{1.2}
\task{$2$}{0.5}{1}{1.2}
\task{$1$}{0.5}{1.5}{1.2}
\task{$p$}{2}{3.5}{1.2}
\task{$x$}{0.5}{4}{1.2}

\task{$3$}{0.5}{0.5}{0.6}
\task{$p$}{2}{2.5}{0.6}
\task{$x$}{0.5}{3}{0.6}
\task{$1$}{0.5}{3.5}{0.6}
\task{$2$}{0.5}{4}{0.6}

\task{$3$}{0.5}{0.5}{0}
\task{$p$}{2}{2.5}{0}
\task{$x$}{0.5}{3}{0}
\task{$2$}{0.5}{3.5}{0}
\task{$1$}{0.5}{4}{0}

\draw[->](0-0.5,0)--(4+0.2,0);
\draw (0, 0.1)--(0,-0.1);

\node[text width=1cm] at (-0.4,2.1) {$101$};
\node[text width=1cm] at (-0.4,1.50) {$101$};
\node[text width=1cm] at (-0.4,0.95) {$99$};
\node[text width=1cm] at (-0.4,0.40) {$99$};



%
%
%

\draw[->](4.2,1.5)--(5.8,1.5);

\draw[->](6-0.2,0)--(9.5+0.1,0);
\draw (6, 0.1)--(6,-0.1);

\task{$x$}{0.5}{6.5}{1.8}
\task{$2$}{0.5}{7}{1.8}
\task{$1$}{0.5}{7.5}{1.8}
\task{$p$}{0.5}{8}{1.8}
\task{$3$}{0.5}{8.5}{1.8}

\task{$3$}{0.5}{6.5}{1.2}
\task{$2$}{0.5}{7}{1.2}
\task{$1$}{0.5}{7.5}{1.2}
\task{$p$}{0.5}{8}{1.2}
\task{$x$}{0.5}{8.5}{1.2}

\task{$3$}{0.5}{6.5}{0.6}
\task{$p$}{0.5}{7}{0.6}
\task{$x$}{0.5}{7.5}{0.6}
\task{$1$}{0.5}{8}{0.6}
\task{$2$}{0.5}{8.5}{0.6}

\task{$3$}{0.5}{6.5}{0}
\task{$p$}{0.5}{7}{0}
\task{$x$}{0.5}{7.5}{0}
\task{$2$}{0.5}{8}{0}
\task{$1$}{0.5}{8.5}{0}
\end{tikzpicture}
\label{fig:profile_couter_example_neutrality}
\end{figure}

For the profile on the left, the only schedule $S$ minimizing the absolute deviation is : $3\succ_S p \succ_S x \succ_S 2 \succ_S 1$.
For the profile on the right, the only schedule $S'$ minimizing the absolute deviation is such that: $3\succ_{S'} 2 \succ_{S'} x \succ_{S'} p \succ_{S'} 1$. Task $p$ has a reduced length but it starts later in $S'$ than in $S$:  \sigmaD does not fulfill {length reduction monotonicity}.
\qed
\end{proof}

\subsection{Reinforcement.}

An aggregation rule $r$ fulfills \emph{reinforcement} (also known as \emph{consistency})~\cite{brandt2016handbook} if, when a ranking $R$ is returned by $r$ on two distinct subsets of voters $A$ and $B$, the same ranking $R$ is returned by $r$ on $A\!\cup\!B$. Since the PTA Kemeny rule sums the weighted Kendall tau score among the voters, it fulfills reinforcement.

\begin{prop}
The PTA Kemeny rule fulfills reinforcement.
\label{prop:PTA_Kemeny_reinforcement}
\end{prop}
\begin{proof}
We consider two subsets of voters $V_1$ and $V_2$.
Since the score is obtained by summing the weighted disagreements over all the voters, the score over $V_1 \cup V_2$ is the sum of the score on $V_1$ and the score on $V_2$. Therefore, if a schedule minimizes the PTA Kendall tau score on both $V_1$ and $V_2$, then it will minimize it on the union of the two subsets. 
\end{proof} 
Note that the PTA Kemeny rule fulfills reinforcement and PTA Condorcet consistency, whereas the already known aggregation rules~\cite{pascual2018collective} for the collective schedule problem either fulfill one or the other but not both. 

\subsection{Unanimity.}

Let us now focus on the \emph{unanimity} axiom, a well-known axiom in social choice. This axiom states that if all the voters rank candidate $a$ higher than candidate $b$ then, in the consensus ranking, $a$ should be ranked higher than $b$. 
We take the same definition, replacing ``rank'' by ``schedule'':

\begin{definition}\textbf{\emph{(Unanimity)}}
Let $P$ be a preference profile and $r$ be an aggregation rule. The rule $r$ fulfills \emph{unanimity} iff when task $a$ is scheduled before another task $b$ in all the preferences in $P$, then $a$ is scheduled before $b$ in any solution returned by $r$.
\end{definition}

Note that this axiom is interesting through its link with precedence constraints in scheduling. Indeed, if all the voters schedule a task before another one, it may indicate that there is a dependency between the two tasks (i.e. a task must be scheduled before the other one). A rule which fulfills the unanimity axiom can then infer the precedence constraints from a preference profile. 

In \cite{pascual2018collective}, the authors prove that the \sigmaT rule does not fulfill unanimity (this property is called Pareto efficiency in the paper). Let us now show that the  \sigmaD does not fulfill this property either. 

\begin{prop}
The \sigmaD rule does not fulfill unanimity for any tie-breaking mechanism.
\label{prop:unanimity_sigmad}
\end{prop}

\begin{proof}
Let us consider an instance with 5 tasks $\{a,b,c,d,e\}$ such that $p_a\!=\!p_b\!=\!p_c\!=\!10$, $p_d\!=\!p_e\!=\!1$ and $\!v\!=\!88$ voters. We consider the following preferences.

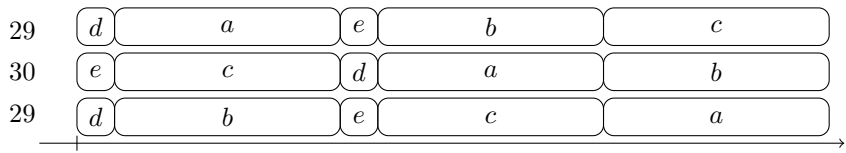
\begin{figure}[H]
\centering
\begin{tikzpicture}

\task{$d$}{0.5}{0.5}{1.2}
\task{$a$}{3}{3.5}{1.2}
\task{$e$}{0.5}{4}{1.2}
\task{$b$}{3}{7}{1.2}
\task{$c$}{3}{10}{1.2}

\task{$e$}{0.5}{0.5}{0.6}
\task{$c$}{3}{3.5}{0.6}
\task{$d$}{0.5}{4}{0.6}
\task{$a$}{3}{7}{0.6}
\task{$b$}{3}{10}{0.6}

\task{$d$}{0.5}{0.5}{0}
\task{$b$}{3}{3.5}{0}
\task{$e$}{0.5}{4}{0}
\task{$c$}{3}{7}{0}
\task{$a$}{3}{10}{0}

\draw[->](0-0.5,0)--(10+0.2,0);
\draw (0, 0.1)--(0,-0.1);

\node[text width=1cm] at (-0.4,1.50) {$29$};
\node[text width=1cm] at (-0.4,0.95) {$30$};
\node[text width=1cm] at (-0.4,0.40) {$29$};

\end{tikzpicture}
\label{fig:profile_couter_example_unanimity_sigmaD}
\end{figure}

In this example, a short task $e$ is always scheduled before a long task $c$ in the preferences. However in the unique optimal solution $S$ for \sigmaD, which is $d \succ_S c \succ_S e \succ_S a \succ_S b$, $e$ is scheduled after $c$. Therefore, the \sigmaD rule does not fulfill unanimity.

Note that, if we reverse all the schedules in the preference profile, then the long task $c$ is always scheduled before the short task $e$ but has to be scheduled after the $e$ in the optimal solution, which is $S$ but reversed.
\end{proof}

One could expect the PTA Kemeny rule to fulfill unanimity since the Kemeny rule does, and since it minimizes the pairwise disagreements with the voters. We can show that this is in fact not the case, by exhibiting a a counter-example.

\begin{prop}
The PTA Kemeny rule does not fulfill unanimity for any tie-breaking mechanism.
\end{prop}

\begin{proof}
We consider an instance with $n=7$ tasks $\{a,b,c,d,e,f,g\}$, such that $p_a=1$, $p_b=10$ and $p_c=p_d=p_e=p_f=p_g=2$, and $v=100$ voters. The preferences are as follows :

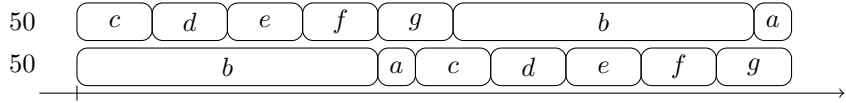
\begin{figure}[H]
\centering
\begin{tikzpicture}
\task{$c$}{1}{1}{0.6}
\task{$d$}{1}{2}{0.6}
\task{$e$}{1}{3}{0.6}
\task{$f$}{1}{4}{0.6}
\task{$g$}{1}{5}{0.6}
\task{$b$}{4}{9}{0.6}
\task{$a$}{0.5}{9.5}{0.6}

\task{$b$}{4}{4}{0}
\task{$a$}{0.5}{4.5}{0}
\task{$c$}{1}{5.5}{0}
\task{$d$}{1}{6.5}{0}
\task{$e$}{1}{7.5}{0}
\task{$f$}{1}{8.5}{0}
\task{$g$}{1}{9.5}{0}

\draw[->](0-0.5,0)--(10+0.2,0);
\draw (0, 0.1)--(0,-0.1);

\node[text width=1cm] at (-0.4,0.95) {$50$};
\node[text width=1cm] at (-0.4,0.40) {$50$};

\end{tikzpicture}
\label{fig:profile_couter_example_unanimity_PTA_Kemeny}
\end{figure}

In this preference profile, the task $b$ is always scheduled before $a$, however in the only optimal solution for PTA Kemeny, $a$ is schedule before $b$, indeed the optimal solution is $a \prec c \prec d \prec e \prec f \prec g \prec b$.
\end{proof}

Note that unanimity is fulfilled if all the tasks are unit tasks. This has indeed already been shown for \sigmaT \cite{pascual2018collective}, this can easily be shown for \sigmaD by using an exchange argument, and this is true for PTA Kemeny since the Kemeny rule fulfills the unanimity axiom. 

In our context, the unanimity axiom is not fulfilled because of the lengths of the tasks. It may indeed be better to disagree with the whole population in order  to minimize the average delay or deviation, for \sigmaT and \sigmaD, or to disagree with the whole population if this disagreement has a small weight, in order to reduce other disagreements which have larger weights, for PTA Kemeny.
Let us now restrict the unanimity axiom to the case where all voters agree to schedule a small task $a$ before a large task $b$: we will see that the solutions returned by the PTA Kemeny rule always schedule $a$ before $b$, that at least one optimal solution returned by \sigmaT also schedules $a$ before $b$, whereas all the optimal solutions for \sigmaD may have to schedule $b$ before $a$ as we can see in the proof of proposition~\ref{prop:unanimity_sigmad}. 

\begin{prop}
Let  $a$ and $b$ be two tasks such that  $p_a \leq p_b$. If task $a$ is always scheduled before task $b$ in the preferences of the voters, then $a$ is scheduled before $b$ in any optimal schedule for the PTA Kemeny rule.
\end{prop}
\begin{proof}
Let us assume, by contradiction,  that a schedule $S$ such that $b$ is scheduled before $a$ is optimal for the PTA Kemeny rule. Let $\mathcal{S}_{(a \leftrightarrow b)}$ be the schedule obtained from $S$ by swapping the position of $a$ and $b$. Let $c$ be a task different from $a$ and $b$. If $c$ is scheduled before $b$ or after $a$ in $S$, then the swap of $a$ and $b$ has no impact on the disagreements with $c$. If $c$ is scheduled between $a$ and $b$, then we have $b \succ_S c$ and $c \succ_S a$ and $a \succ_{\mathcal{S}_{(a \leftrightarrow b)}} c$ and $c\succ_{\mathcal{S}_{(a \leftrightarrow b)}} b$  (the order between $c$ and the tasks other than $a$ and $b$ does not change between $\mathcal{S}_{(a \leftrightarrow b)}$ and $S$). Task $a$ is always scheduled before task $b$ in the preferences and $p_a \leq p_b$, therefore the overall cost of scheduling $a$ before $c$ is smaller than or equal to the cost of scheduling $b$ before $c$. Furthermore, since $a$ is always scheduled before $b$ in the preferences, scheduling $a$ before $b$ does not create a new disagreement, whereas the cost of scheduling $b$ before $a$ is equal to $v \cdot p_b$. The overall cost of $S$ is then strictly larger than the cost of $\mathcal{S}_{(a \leftrightarrow b)}$ which means that $S$ is not optimal, a contradiction.
\end{proof}


\begin{prop}
Let  $a$ and $b$ be two tasks such that $p_a\leq p_b$. If task $a$ is always scheduled before task $b$ in the preferences of the voters, then $a$ is scheduled before $b$ in at least one optimal schedule for the \sigmaT rule.
\end{prop}

\begin{proof}
Suppose that an schedule $S$ is optimal for the \sigmaD rule and such that $b$ is scheduled before $a$ in $S$. By swapping the positions of $a$ and $b$ in $S$, we obtain a new feasible solution $S$ in which the completion times of all tasks but $a$ and $b$ are is either smaller than or equal to the ones in $S$. The completion time of $b$ in $S'$ is the one of $a$ in $S$ an the completion time of $a$ in $S'$ is smaller than or equal to the one of $b$ in $S$. The completion time of $b$ in each preference is strictly higher than the completion time of $a$. Therefore, in $S'$ the tardiness of $b$ which is ending in $S'$ at the time $a$ ends in $S$, is smaller than or equal to the tardiness of $a$ in $S$, similarly the tardiness of $a$ in $S'$ is smaller than or equal to the tardiness of $b$ in $S$. Overall, the tardiness of $S'$ is smaller than or equal to the tardiness of $S$. Their tardiness are equal if both $a$ and $b$ are scheduled before their minimum completion time in the preference profile.
\end{proof}

Thus, if we are looking for a single solution for \sigmaT, we can restrict the search to solutions fulfilling the unanimity axiom for couples of tasks for which all the voters agree that the smaller task should be scheduled first. For \sigmaD, we can guarantee solutions which fulfill this axiom for couples of tasks of the \nolinebreak same \nolinebreak length. 

\begin{prop}
Let  $a$ and $b$ be two tasks such that $p_a=p_b$. If task $a$ is always scheduled before task $b$ in the preferences of the voters, then $a$ is scheduled before $b$ in at least one optimal schedule for the \sigmaD rule.
\label{prop:unanimity_egalite_sigmad}
\end{prop}

\begin{proof}
Suppose that an schedule $S$ is optimal for the \sigmaD rule and such that $b$ is scheduled before $a$ in $S$. By swapping the positions of $a$ and $b$ in $S$, we obtain a new feasible solution $S_{(a \leftrightarrow b)}$ in which the completion times of all tasks but $a$ and $b$ are is either lower or equal to the ones in $S$. The completion time of $b$ in $S_{(a \leftrightarrow b)}$ is the one of $a$ in $S$ an the completion time of $a$ in $S_{(a \leftrightarrow b)}$ is lower or equal to the one of $b$ in $S$. The completion time of $b$ in each preference is strictly higher than the completion time of $a$. Therefore, in $S_{(a \leftrightarrow b)}$ the tardiness of $b$ which is ending in $S_{(a \leftrightarrow b)}$ at the time $a$ ends in $S$, is lower or equal to the tardiness of $a$ in $S$, similarly the tardiness of $a$ in $S_{(a \leftrightarrow b)}$ is lower or equal to the tardiness of $b$ in $S$. Overall, the tardiness of $S_{(a \leftrightarrow b)}$ is lower or equal to the tardiness of $S$. Their tardiness are equal if both $a$ and $b$ are scheduled before their minimum completion time in the preference profile.
\end{proof}

We have seen that for the \sigmaT and PTA Kemeny rules, if a task $a$ is scheduled before a task $b$ and $a$ is not longer than $b$, then there exists an optimal solution which schedules $a$ before $b$. This is not the case for \sigmaD. In EB settings, we would expect well supported short tasks to be scheduled before less supported large tasks. Therefore the \sigmaT and PTA Kemeny rules seem well adapted for EB settings, while the \sigmaD rule seems less relevant in these settings.

\subsection{Summary.}
We summarize the results shown in this section in Table~\ref{tab:axioms} (a ``*'' means that the property has been showed in \cite{pascual2018collective}, the other results are shown in this paper).
\begin{table}[H]
\centering
\footnotesize
\begin{tabular}{|c|c|c|c|c|c|c|c|c|c|}
 \hline
 &&&&&& & \multicolumn{3}{c|}{Unanimity \scriptsize{($a$ before $b$)}}\\
 Rule & N & PTA N & R & PTA C & LRM & Distance & $\!p_a\!<\!p_b\!$ & $\!p_a\!=\!p_b\!$ & $\!p_a\!>\!p_b\!$\\
 \hline
 PTA K & {\color{red} \xmark} &  {\color{green} \cmark} & {\color{green} \cmark} & {\color{green} \cmark} & ? & {\color{red} \xmark} & {\color{green} \cmark} &  {\color{green} \cmark} & {\color{red} \xmark}\\
 \sigmaT & {\color{red} \xmark} &  {\color{green} \cmark} & {\color{green} \cmark}* & {\color{red} \xmark}* & ? & {\color{red} \xmark} & {\color{orange} $\sim$} &  {\color{orange} $\sim$} & {\color{red} \xmark}*\\
 \sigmaD & {\color{red} \xmark}  & {\color{green} \cmark} & {\color{green} \cmark}* & {\color{red} \xmark} & {\color{red} \xmark} & {\color{green} \cmark} & {\color{red} \xmark}  & {\color{orange} $\sim$} & {\color{red} \xmark}\\
  \hline
\end{tabular}
\smallskip
\caption{Fulfilled ({\color{green} \cmark}) and unfulfilled ({\color{red} \xmark}) axioms by the PTA Kemeny, \sigmaT and \sigmaD rules. Symbol {\color{orange} $\sim$}  means that the property is fulfilled by at least one optimal solution. The acronyms in the columns correspond to: neutrality (N), PTA neutrality (PTA N), reinforcement (R), PTA Condorcet consistency (PTA C), length reduction monotonicity (LRM).} 
\label{tab:axioms}
\end{table}

\section{Computational complexity and algorithms}
\label{sec:resolution}

In this section we  study the computational complexities of the \sigmaD and the PTA Kemeny rules. We will then focus on resolution methods for these rules. 
The \sigmaT rule has already been proven to be {strongly} NP-hard~\cite{pascual2018collective}. In the same work, authors use linear programming to solve instances up to 20 tasks and 5000 voters, which is satisfactory since realistic instances are likely to have few tasks and a lot of voters. 

The PTA Kemeny rule is NP-hard to compute since it is an extension of the Kemeny rule, which is NP-hard to compute~\cite{bartholdi1989voting}. Most of the algorithms used to compute the ranking returned by the Kemeny rule can be adapted to return the schedule returned by the PTA Kemeny rule, by adding weights on the disagreements in the resolution method.
In the following section, to compute schedules returned by the PTA Kemeny rule, we will use a weighted  adaptation of an exact linear programming formulation for the Kemeny rule \cite{conitzer2006improved}. 

Regarding the \sigmaD rule, when there are exactly two voters, the problem is easy to solve: we return one of the two schedules in the preference profile (since deviation is a distance, any other schedule would have a larger  deviation to the profile because of triangle inequalities). 
In the general case, the problem is NP-hard, as shown below. 

\begin{theorem}
The problem of returning a schedule minimizing the total absolute deviation is strongly NP-hard.
\label{thm:np_hard}
\end{theorem}

In order to prove that computing the schedules returned by \sigmaD is NP-hard, we introduce and prove one preliminary lemma.
In the sequel, for reasons of readability, we will denote by \sigmaDp the problem which consists in returning a schedule which minimizes the sum of the absolute deviations with the preference profile (i.e. \sigmaDp is the problem solved by the \sigmaD rule).   

In the sequel, we consider a polynomial time reduction from the problem $(1|no-idle|\Sigma D)$ which has been proven as strongly NP-hard when coded in unary \cite{wan2013single}. In this problem we consider an instance $I$ composed of a set $J$ of $n$ tasks, each task $i$ having a processing time $p_i \in \mathbb{N}^*$ and a deadline $d_i$. By  $L=\sum_{i\in J} p_i$, we denote the overall load of the tasks. A feasible solution for this problem is a schedule $S$ of all tasks in $J$ on a single machine, with no idle time. We denote by $D(S)$ the sum of the absolute deviations $ \sum_{i \in J} |C_i(S)-d_i|$ where $C_i(S)$ is the completion time of task $i$ in the schedule $S$. Given an integer $B$, the objective is to determine if a schedule $S$ with a total sum of absolute deviations $D(S)$ smaller than $B$ exists. Since no task can complete before its processing time or after $L$ (because there is no idle time), we can assume without loss of generality that $L \geq d_i \geq p_i$. If a task $i$ had a deadline smaller than its processing time, then all feasible solutions would at least have a deviation of $p_i-d_i$ for task $i$, therefore, we can reduce $B$ by that amount and have $d_i=p_i$; an analogous remark can be done for $d_i > L$.\\

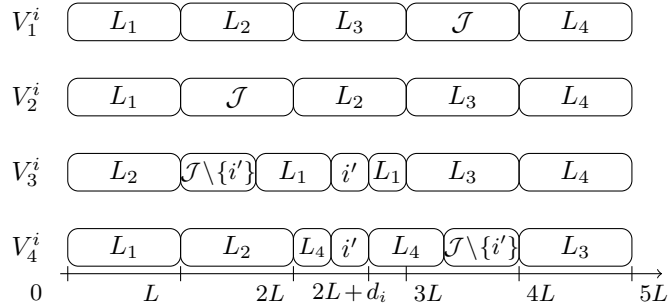
\begin{figure}[H]
\centering
\begin{tikzpicture}
\task{$L_1$}{1.5}{1.5}{3}
\task{$L_2$}{1.5}{3}{3}
\task{$L_3$}{1.5}{4.5}{3}
\task{$\mathcal{J}$}{1.5}{6}{3}
\task{$L_4$}{1.5}{7.5}{3}

\task{$L_1$}{1.5}{1.5}{2}
\task{$\mathcal{J}$}{1.5}{3}{2}
\task{$L_2$}{1.5}{4.5}{2}
\task{$L_3$}{1.5}{6}{2}
\task{$L_4$}{1.5}{7.5}{2}

\task{$L_2$}{1.5}{1.5}{1}
\task{\small $\mathcal{J}\!\setminus\!\{i'\}$}{1}{2.5}{1}
\task{$L_1$}{1}{3.5}{1}
\task{$i'$}{0.5}{4}{1}
\task{$L_1$}{0.5}{4.5}{1}
\task{$L_3$}{1.5}{6}{1}
\task{$L_4$}{1.5}{7.5}{1}

\task{$L_1$}{1.5}{1.5}{0}
\task{$L_2$}{1.5}{3}{0}
\task{\small $L_4$}{0.5}{3.5}{0}
\task{$i'$}{0.5}{4}{0}
\task{$L_4$}{1}{5}{0}
\task{\small $\mathcal{J}\!\setminus\!\{i'\}$}{1}{6}{0}
\task{$L_3$}{1.5}{7.5}{0}

\draw[->](0-0.1,0)--(7.5+0.4,0);
\draw (0, 0.1)--(0,-0.1);
\draw (4, 0.1)--(4,-0.1);
\draw (1.5, 0.1)--(1.5,-0.1);
\draw (3, 0.1)--(3,-0.1);
\draw (4.5, 0.1)--(4.5,-0.1);
\draw (6, 0.1)--(6,-0.1);
\draw (7.5, 0.1)--(7.5,-0.1);

\node[text width=1cm] at (0,-0.25) {\small $0$};
\node[text width=1cm] at (1.5,-0.25) {\small $L$};
\node[text width=1cm] at (3,-0.25) {\small $2L$};
\node[text width=1cm] at (3.75,-0.25) {\small $2L+d_i$};
\node[text width=1cm] at (5.1,-0.25) {\small $3L$};
\node[text width=1cm] at (6.6,-0.25) {\small $4L$};
\node[text width=1cm] at (8.1,-0.25) {\small $5L$};

\node[text width=1cm] at (-0.25,3.35) {$V^i_1$};
\node[text width=1cm] at (-0.25,2.35) {$V^i_2$};
\node[text width=1cm] at (-0.25,1.35) {$V^i_3$};
\node[text width=1cm] at (-0.25,0.35) {$V^i_4$};

\end{tikzpicture}

\caption{Voters associated with a task $i'$}
\label{fig:profil_reduction}
\end{figure}

From an instance $I$ of the $(1|no-idle|\Sigma D)$ problem, we define an instance $I'$ of the \sigmaDp problem. We have a set $\mathcal{J}$ of $n+4L$ tasks. For each task $i$ of $J$, there is a task $i'$ in $\mathcal{J}$, with $p_{i'}=p_i$. The set $\mathcal{J}$ also contains $4L$ tasks of length $1$. These $4L$ tasks are partitioned into 4 sets $L_1$, $L_2$, $L_3$ and $L_4$, all these tasks are of length 1. Instance $I'$ has $4n$ voters: for each task $i \in J$, we create 4 voters $V^i_1,V^i_2,V^i_3,V^i_4$ (see Figure~\ref{fig:profil_reduction}). 

\begin{itemize}
	\item Voter $V^i_1$, of \emph{type 1}, schedules first the tasks of $L_1$ , then the tasks of $L_2$, then the tasks of $L_3$, then the tasks of $\mathcal{J}$ and finally the tasks of $L_4$.
	\item Voter $V^i_2$, of \emph{type 2}, schedules firstly the tasks of $L_1$, followed by the tasks of $\mathcal{J}$, then the tasks of $L_2$, then the tasks of $L_3$ and finally the tasks of $L_4$.
	\item Voter $V^i_3$, of \emph{type 3}, schedules task $i'$ in order that this task completes at time $d_i+2L$. The rest of the schedule is as follows: first, the tasks of $L_2$, then the tasks of $\mathcal{J}$ except $i'$, then the tasks of $L_1$ are scheduled around task $i'$. The schedule ends with the tasks of $L_3$ followed by the tasks of $L_4$.
	\item Voter $V^i_4$, of \emph{type 4}, schedules task $i'$ in order that this task completes at time $d_i+2L$. The rest of the schedule is as follows: first, the tasks of $L_1$, then the tasks of $L_2$, then the tasks of $L_4$ are scheduled around task $i'$, then the tasks of $\mathcal{J}$ without $i'$. The tasks of $L_3$ end the schedule.
\end{itemize}

The order on the tasks in each of the subsets $L_1$, $L_2$, $L_3$, $L_4$ is the same for all voters. For the set $\mathcal{J}$, the order is the same for all voters, but, for each voter of type 3 and 4 one task is scheduled at a given time, regardless of its usual rank in the order.

Let us note that we can create such an instance in polynomial time since the instance for the $(1|no-idle|\sum D)$ problem is coded in unary.

\begin{customlemma}{1}
Given an instance $I$ of the $(1|no-idle|\sum D)$ problem, there exists an optimal solution for the instance $I'$ of \sigmaDp, created as described above, in which the tasks are scheduled as follows: $L_1$ first, $L_2$ second, $\mathcal{J}$ third, then $L_3$ and finally $L_4$.
\label{lem:structure_sol}
\end{customlemma}

\begin{proof}
\emph{Fact 1: There exists an optimal schedule $S^*$ with tasks of $L_1$ scheduled before tasks of $L_2$ and $\mathcal{J}$.}\\
To 
prove \emph{fact 1}, we consider an optimal schedule $S^*$ in which at least one task of $L_1$ is scheduled after a task of $\mathcal{J}$ or $L_2$. Thanks to proposition~\ref{prop:unanimity_egalite_sigmad}, we can consider that in $S^*$, the tasks of $L_1$ and $L_2$ are scheduled before the tasks of $L_3$ and $L_4$ and in the same order than in the preferences.  Let us call $x_1 \in L_1$ the first task of $L_1$ scheduled just after a task $x_2 \in L_2 \cup \mathcal{J}$. We note $C_{x_1}(S^*)$ and $C_{x_2}(S^*)$ the completion times of tasks $x_1$ and $x_2$ in $S^*$. Since $x_1$ is the first task of $L_1$ to be scheduled just after a task of $L_2$ or $\mathcal{J}$ and since the tasks of $L_1$ are scheduled in the same order as in the preferences, task $x_2$ starts after the tasks of $L_1$ preceding $x_1$ in the preferences. We study the schedule $S$ in which the tasks $x_1$ and $x_2$ are swapped. We distinguish two cases:
\begin{enumerate}
	\item Task $x_2$ is in $L_2$: the swap changes the order on the tasks $x_1$ and $x_2$, the order on the tasks of $L_1$ (resp. $L_2$) is unchanged. Therefore $x_1$ (resp. $x_2$) is still scheduled after (resp. before) the tasks scheduled after it (resp. before it) in the preferences, which means that, in $S$, the task $l$ (resp. $j$) completes at or after (resp. at or before) its completion time for voters of type 1,2 and 4 (resp. for voters of type 2). Thus, for each voter of type 1 and 4, the absolute deviation is reduced by one, and reduced by two for each voter of type 2. Overall, the absolute deviation is reduced by $4n$. On the other hand, voters of type 1,3 and 4 could increase their deviations of 1 relatively to task $x_2$ and voters of type 3 could also increase their deviation for the task $x_1$ of 1. In the worst case, this increase is of $4n$, which equals the reduction, therefore $S$ would also be optimal.
	\item Task $x_2$ is in $\mathcal{J}$: following the same reasoning, we can see that voters of type 1,2 and 4 will decrease their deviations for task $x_1$ by $p_{x_2}$ with the swap. Voters of type 1 will also decrease their deviation for task $x_2$ by one since $x_1$ completes before $3L$ in $S^*$ which implies that $x_2$ completes before $3L$ in $S$. Overall the reduction is of $3np_j+n$. Voters of type 2,3 and 4 could increase their deviation for $j$ by 1 and voters of type 3 could increase their deviation with $l$ by $p_i$, overall the increase is at most of $3n+np_i$, since $p_i \geq 1$ the increase is smaller than or equal to the decrease, $S$ is also optimal.
\end{enumerate}
In both cases, $S$ is at least as good as $S^*$, therefore, from any optimal solution respecting proposition~\ref{prop:unanimity_egalite_sigmad}, we can iteratively obtain a new optimal solution in which the tasks of $L_1$ are scheduled before the tasks of $L_2$ and $\mathcal{J}$.\\

\emph{Fact 2: In $S^*$, tasks of $L_4$ are scheduled after tasks of $L_3$ and $\mathcal{J}$.}\\
We can prove \emph{fact 2} in an analogous way than \emph{fact 1}, but symmetrically.\\

\emph{Fact 3: In $S^*$, tasks of $L_2$ are scheduled before tasks of $\mathcal{J}$.}\\
We show that there is an optimal solution in which the tasks of $L_2$ are scheduled before the tasks of $\mathcal{J}$. We consider an optimal solution $S^*$, respecting the previous facts and proposition~\ref{prop:unanimity_egalite_sigmad}, in which at least one task of $L_2$ is scheduled after a task of $\mathcal{J}$. Let us denote by $l$ the first task of $L_2$ scheduled after a task of $\mathcal{J}$. Let us call $j$ the task of $\mathcal{J}$ scheduled just before $l$ in $S^*$. Note that such  a task always exists since the task of $L_1$ are scheduled before the tasks of $\mathcal{J}$ and $L_2$ and the tasks of $L_3$ and $L_4$ are scheduled after the ones of $L_2$. We study the schedule $S$, similar to $S^*$ except that $l$ and $j$ are swapped. Since the tasks of $L_2$ are in the same order than in the preferences, and since we swap $l$ only with tasks of $\mathcal{J}$, $l$ cannot complete in $S$ before tasks of $L_2$ scheduled before it in $S^*$. Therefore, $l$ completes in $S$ at least at the same time than in the preferences of voters of type 1 and 4. By swapping $l$ and $j$, we reduce the absolute deviation on $l$ for voters of type 1,3 and 4 by $p_j$, we also reduce absolute deviation on $j$ for voters of type 1, by 1. Overall, we reduce the sum of absolute deviation by $3np_i+n$. We may increase the absolute deviation on $j$ for voters of type 2,3 and 4 by one and deviation on $l$ for voters of type 2 by $p_i$, increasing the total sum of deviation by at most $3n+np_i$. Since $p_i \geq 1$, the increase is smaller than the decrease, therefore $S$ is also  optimal.\\

\emph{Fact 4: In $S^*$, tasks of $L_3$ are scheduled after tasks of $\mathcal{J}$.}\\
We can prove \emph{fact 4} in the same way than \emph{fact 3}.\\

From facts 1 to 4, we get Lemma~\ref{lem:structure_sol}.

\begin{figure}[H]
\centering
\begin{tikzpicture}
\task{$L_1$}{1.5}{1.5}{0}
\task{$L_2$}{1.5}{3}{0}
\task{$J'$}{1.5}{4.5}{0}
\task{$L_3$}{1.5}{6}{0}
\task{$L_4$}{1.5}{7.5}{0}

\draw[->](0-0.1,0)--(7.5+0.4,0);
\draw (0, 0.1)--(0,-0.1) node[below]{$0$};
\draw (1.5, 0.1)--(1.5,-0.1) node[below]{$L$};
\draw (3, 0.1)--(3,-0.1) node[below]{$2L$};
\draw (4.5, 0.1)--(4.5,-0.1) node[below]{$3L$};
\draw (6, 0.1)--(6,-0.1) node[below]{$4L$};
\draw (7.5, 0.1)--(7.5,-0.1) node[below]{$5L$};

\end{tikzpicture}
\caption{Structure of an existing optimal solution}
\end{figure}
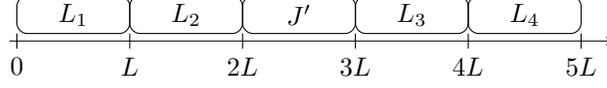
\end{proof}

We can now prove that the \sigmaD rule solves an NP-hard problem.

\begin{customthm}{1}
The problem of returning a schedule minimizing the total absolute deviation (\sigmaDp) is strongly NP-hard.
\end{customthm}

\begin{proof}
In a schedule which follows the structure explained in Lemma~\ref{lem:structure_sol}, it is possible to calculate the absolute deviations associated with the tasks of subsets $L_1$, $L_2$, $L_3$ and $L_4$:
\begin{itemize}
	\item $L_1$: the tasks of $L_1$ are scheduled exactly like in the preference of voters of type 1, 2 and 4. The voter of type 3 associated with the task $i$ has a delay of $L+(L-p_i)$ on the $d_i$ first tasks of $L_1$ and a delay of $2L$ on the others. Overall, the deviation is:  $n \times \sum_i d_i \times(2L-p_i)+(L-d_i)\times 2L$.
	\item $L_2$: Voters of type 1 and 4 have no deviation on the tasks of $L_2$. The $2n$ voters of type 2 and 3 have a deviation of $L$ on each of the $L$ tasks of $L_2$, which amounts to $2nL \times L$.
	\item $L_3$: Symmetrically to $L_2$, the sum of deviation of tasks of $L_3$ is also: $2nL \times L$.
	\item $L_4$: Voters of type 1,2 and 3 have no deviation on the tasks of $L_4$. For the voters of type 4: the first $d_i-p_i$ tasks of $L_4$ are delayed of $2L$, the rest of the tasks of $L_4$ are delayed by $2L-p_i$, which amounts to: $n \times \sum_i (d_i-p_i) \times 2L +(L-(d_i-p_i)) \times(2L - p_i)=n \times \sum_i (d_i-p_i)(p_i)+2L^2-Lp_i$.
\end{itemize}
Overall the sum of deviations $M$ associated with the subsets $L_1$, $L_2$, $L_3$ and $L_4$ is:
\[
M=\left(n \times \sum_i d_i \times(2L-p_i)+(L-d_i)\times 2L\right)
\]
\[
+\left( 2nL^2 \right) +\left( 2nL^2\right) +\left(n \times \sum_i (d_i-p_i)(p_i)+2L^2-Lp_i \right)
\]

\[
M\!=\!4nL^2+ \left(n \sum_i -p_id_i + 2L^2 +p_id_i-p_i^2+2L^2-Lp_i \right)
\]
\[
M=4nL^2 +n \times \sum_i 4L^2-p_i(L+p_i)
\]

We now study the deviation of the tasks of $\mathcal{J}$. The median completion time of task $i'$ in $\mathcal{J}$ is $d_{i'}=2L+d_i$. Let us see that, regardless of the order on the tasks of $\mathcal{J}$ in the preference, voters of type 1 and 2 will always have a total deviation on task $i'$ of 2L. Since the order is the same for all the voters, the task will complete at a time $L+K$ with $K$ an integer lower than $L$, in the preference of any voter of type 2 and at $3L+K$ in the preference of any voter of type 1. Therefore, in any schedule $S$, since the task completes at a time $C_{i'}(S)$ between $2L$ and $3L$, we will have a total deviation of $C_{i'}(S)-(L+K) + (3L+K)-C_{i'}(S)=2L$. We count then $2L$ for every pair of voter of type 1 and 2, which amounts to $2L \times n$ for each task, so the overall deviation of $2Ln^2$.\\

For the last two type of voters, for each task $i'$, we distinguish two cases:
\begin{enumerate}
	\item Voters $V^i_3$ and $V^i_4$ both have scheduled $i'$ so it completes at $2L+d_i=d_{i'}$. The deviation of a schedule $S$, regarding these two voters is therefore $2\lvert C_{i'}(S)-d_{i'} \rvert$.
	\item All other voters of type 3 and 4 schedule tasks of $\mathcal{J}$ in the same order except of one task $j'$, which is scheduled to complete at $d_{j'}$. Let us denote by $K$ the integer such that task $i'$ completes at $L+K$ in $V_3^j$, then $i'$ completes at $3L+p_{j'}+K$ in $V_4^j$. Since $i'$ completes between $2L$ and $3L$ in the optimal solution $S$ we are considering, we know that the deviation with $V_3^j$ and $V_4^j$ regarding task $i'$ will be $C_{i'}(S)-(L+K)+3L+p_{j'}+K-C_{i'}(S)=2L+p_{j'}$. We calculate this value for all tasks, and call it $N$: 
\[
N=\sum_{i' \in \mathcal{J}} \left(\sum_{j' \in \mathcal{J} \setminus \{i'\}} 2L+p_{j'} \right)
\]
\[
N=\sum_{i' \in \mathcal{J}} 2L(n-1)+L-p_{i'}=2Ln^2-Ln-L
\]
\end{enumerate} 
By summing all these terms, the deviation of a solution $S'$ respecting lemma~\ref{lem:structure_sol} is: 
\[
D(S')=M+2Ln^2+N +2\sum_{i' \in \mathcal{J}} \lvert C_{i'}(S')-d_{i'} \rvert
\]
If a solution $S$ with a cost lower than $B$ exists for instance $I$ of problem $(1|no-idle|\Sigma D)$, then there is a solution $S'$ with a cost lower than $M+2Ln^2+N+2B$ for instance $I'$ of \sigmaDp. We can find this solution by reproducing the order on the task of $J$ on the tasks on $\mathcal{J}$. More precisely, $S'$ respects Lemma~\ref{lem:structure_sol} and schedules task from $\mathcal{J}$ in the order corresponding to $S$ with the tasks of $J$. We would have $C_{i'}(S')=C_i(S)+2L$ and $d_i'=d_i+2L$. Therefore, for all $i$, we have $\lvert C_{i}(S)-d_{i} \rvert=\lvert C_{i'}(S')-d_{i'} \rvert$. Since, $\sum_{i \in J} \lvert C_{i}(S)-d_{i} \rvert \leq B$, we have $\sum_{i' \in \mathcal{J}} \lvert C_{i'}(S')-d_{i'} \rvert \leq B$ and consequently, $D(S') \leq M+2Ln^2+N+2B$.\\
Reciprocally, if there exists a solution $S'$ with a total deviation $D(S')$ smaller than $M+2Ln^2+N+2B$ for an instance $I'$ of \sigmaDp, we can create a solution $S$ with a cost lower than $B$ for an instance $I$ of $(1|no-idle|\sum D)$ by recreating the order on the tasks of $\mathcal{J}$ on the tasks of $J$.\\

We showed that there exists a solution of cost at most $B$ for the $(1|no-idle|\Sigma D)$ problem for instance $I$ iff there is a solution of cost at most $M+2Ln^2+N+2B$ for instance $I'$ of \sigmaDp, that we can obtain in polynomial time. Since $(1|no-idle|\Sigma D)$ is strongly NP-hard, \sigmaDp is strongly NP-hard.
\end{proof}

Since computing an optimal schedule for \sigmaD is strongly NP-hard, we propose two resolution methods.  First, we use linear programming, allowing us to solve exactly instances up to 15 tasks in less than 30 minutes. Second, we propose a heuristic and the use of local search to improve the solution of the heuristic. 

\medskip

\noindent \textbf{A heuristic for \sigmaD: LMT.}
The heuristic we propose is called LMT, which stands for ``Lowest Median Time''. For each task of $\mathcal{J}$, we compute its median completion time in the preferred schedules of the voters. The LMT algorithm then consists in scheduling the tasks by non decreasing median completion times. 

The idea behind LMT is the following one: the closer the completion time of a task  is to its median completion time, the lower is its deviation. 
As we will see in Section~\ref{sec:experiments}, LMT performs well in practice, even if, in the worst cases, it can lead to really unsatisfactory schedules, which can be shown by exhibiting a worst case instance. 

\begin{prop}
For any $\alpha \geq 1$, LMT is not $\alpha$-approximate for the total absolute deviation minimization.
\end{prop}


\begin{proof}
Let us consider an instance with $v$ voters and $n$ tasks. Tasks $t_1$, $t_2$ and $t_3$ are of size $p$, with $p$ an integer and $n-3$ tasks $t_4, \dots, t_n$ are of size $1$. We consider the following preference profile:

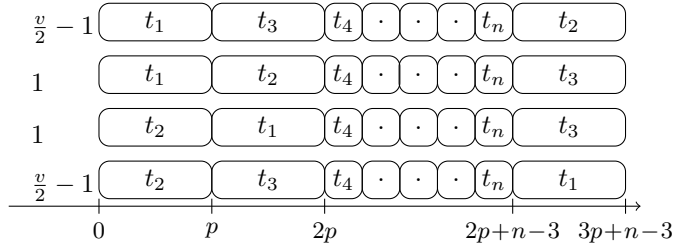
\begin{figure}[H]
\centering
\begin{tikzpicture}
\task{$t_1$}{1.5}{1.5}{2.1}
\task{$t_3$}{1.5}{3}{2.1}
\task{$t_4$}{0.5}{3.5}{2.1}
\task{$.$}{0.5}{4}{2.1}
\task{$.$}{0.5}{4.5}{2.1}
\task{$.$}{0.5}{5}{2.1}
\task{$t_n$}{0.5}{5.5}{2.1}
\task{$t_2$}{1.5}{7}{2.1}

\task{$t_1$}{1.5}{1.5}{1.4}
\task{$t_2$}{1.5}{3}{1.4}
\task{$t_4$}{0.5}{3.5}{1.4}
\task{$.$}{0.5}{4}{1.4}
\task{$.$}{0.51}{4.5}{1.4}
\task{$.$}{0.5}{5}{1.4}
\task{$t_n$}{0.5}{5.5}{1.4}
\task{$t_3$}{1.5}{7}{1.4}

\task{$t_2$}{1.5}{1.5}{0.7}
\task{$t_1$}{1.5}{3}{0.7}
\task{$t_4$}{0.5}{3.5}{0.7}
\task{$.$}{0.5}{4}{0.7}
\task{$.$}{0.5}{4.5}{0.7}
\task{$.$}{0.5}{5}{0.7}
\task{$t_n$}{0.5}{5.5}{0.7}
\task{$t_3$}{1.5}{7}{0.7}

\task{$t_2$}{1.5}{1.5}{0}
\task{$t_3$}{1.5}{3}{0}
\task{$t_4$}{0.5}{3.5}{0}
\task{$.$}{0.5}{4}{0}
\task{$.$}{0.5}{4.5}{0}
\task{$.$}{0.5}{5}{0}
\task{$t_n$}{0.5}{5.5}{0}
\task{$t_1$}{1.5}{7}{0}

\draw[->](0-1.2,0)--(7+0.2,0);

\node[text width=1cm] at (-0.4,2.35) {$\frac{v}{2}-1$};
\node[text width=1cm] at (-0.4,1.65) {$1$};
\node[text width=1cm] at (-0.4,0.95) {$1$};
\node[text width=1cm] at (-0.4,0.25) {$\frac{v}{2}-1$};

\draw (0, 0.1)--(0,-0.1) node[below]{\small{$0$}};
\draw (1.5, 0.1)--(1.5,-0.1) node[below]{\small{$p$}};
\draw (3, 0.1)--(3,-0.1) node[below]{\small{$2p$}};
\draw (5.5, 0.1)--(5.5,-0.1) node[below]{\small{$2p\!+\!n\!-\!3$}};
\draw (7, 0.1)--(7,-0.1) node[below]{\small{$3p\!+\!n\!-\!3$}};
\end{tikzpicture}
\end{figure}

In such an instance, tasks $t_1$ and $t_2$ have median completion times $m_1=m_2=p$. The task $t_3$ has a median completion time $m_3=2p$. Tasks $t_4$ to $t_n$ have median completion times from $2p+1$ to $2p+n-3$. Therefore LMT returns a schedule with $t_1$ and $t_2$ first, in any order, then $t_3$ and finally $t_4$ to $t_n$ in this order. This solution has a sum of deviation of $\sum D =vpn+vn-3v-4p$.

Let us now consider another solution: $t_1 \succ t_3 \succ t_4 \succ ... \succ t_n \succ t_2$, we calculate its total deviation and find: $\sum D = 2pv+vn-3v+2p+2n-6$.

We calculate the ratio between the two values:
\[
\frac{vpn+vn-3v-4p}{2pv+vn-3v+2p+2n-6}
\]
When $p,n$ and $v$ tend towards $+\infty$ the ratio tends towards $+\infty$ as well. Therefore the LMT algorithm can return a schedule with a sum of deviations arbitrarily far from the optimal one.
\end{proof}


\noindent \textbf{Local search.}
In order to improve the solution returned by our heuristic, we propose a local search algorithm. We define the neighbourhood of a schedule $S$ as the set of schedules obtained from $S$  in which two consecutive tasks have been swapped. 
If at least one neighbour has a total deviation smaller than $S$, we choose the best one and we restart from it. Otherwise, $S$ is a local optimum and we stop the algorithm.
At each step, 
we study $(n-1)$ neighbours: the complexity is linear with the number of steps. 
In our experiments,  
by  letting the algorithm reach a local optimum, we saw that the result obtained is usually very close to its local optimum at $n$ steps and, that the local search always ends before $2n$ steps: in practice, we can bound the number of steps to $2n$ without reducing the quality of the solution. 


\vspace{-1em}
\section{Experiments}
\label{sec:experiments}
\vspace{-0.5em}

\noindent\textbf{Instances.}
Since no database of instances for the collective schedules problem exists, we use synthetic instances. We generate two types of preference profiles: uniform (denoted below by U), in which the  preferences are drawn randomly, 
and correlated (C) in which the preferences are drawn according to the Plackett-Luce model \cite{plackett1975analysis,luce2012individual}. In this model, each task $i$ has an objective utility $u_i$ (the utilities of the tasks are drawn uniformly in the [0,1] interval). We consider that the voters pick the tasks sequentially (i.e. they choose the first task of the schedule, then the second, and so forth). When choosing a task in a subset $J$, each task $i$ of $J$ has a probability of being picked of $u_i/\sum_{j \in J} u_j$.
The lengths of the tasks are chosen uniformly at random between 1 and 10 (the results do not differ when the  lengths are chosen in interval [1,5]). 
For all the experiments, we will use linear programming (CPLEX) to compute one optimal solution for each rule. Note that for most of the instances we generated, our rules had only one optimal solution. This was the case for more than $99\%$ of the instances for \sigmaT and \sigmaD. For PTA Kemeny, this was the case for  about $90\%$ (resp. $95\%$) of the  instances for PTA Kemeny when the instance had 100 voters (resp. 250 voters), and for $98\%$ of cases in correlated instances with 250 voters. 

\medskip
\noindent\textbf{Computation times.}
We run the two linear programming algorithms corresponding to the \sigmaD and PTA Kemeny rules. The experiments are run on a 6-core Intel i5 processor. The mean computation times can be found in Table~\ref{tab:computation_times}.

\begin{table}[H]
    \centering
    \begin{tabular}{c|c||c|c|c||c|c|c}
        \multicolumn{2}{c||}{} & \multicolumn{3}{c||}{\small \sigmaD} &  \multicolumn{3}{c}{\small PTA Kemeny}\\
        Nb voters & $P$ & $n\!=\!4$ & $n\!=\!8$ & $n\!=\!12$  & $n\!=\!4$ & $n\!=\!8$ & $n\!=\!12$\\
        \hline
        \multirow{2}{*}{$50$} & U & 0.01 & 0.28 & 10.4 & 0.004 & 0.02 & 0.05\\
         & C & 0.005 & 0.13 & 0.95 & 0.002 & 0.02 & 0.05\\
        \hline
         \multirow{2}{*}{$500$} & U & 0.01 & 25.0 & 104.1 & 0.003 & 2.1 & 4.6\\
         & C & 0.006 & 13.4 & 47.6 & 0.003 & 1.3 & 3.8\\
         \hline
    \end{tabular}
    \caption{Mean computation times (s) for \sigmaD and PTA Kemeny.}
    \label{tab:computation_times}
\end{table}


These algorithms allow to solve small but realistic instances. 
Note that correlated instances, which are more likely to appear in realistic settings, require less computation time than uniform ones. Note also that computing an optimal schedule for PTA Kemeny is way faster than an optimal schedule for \sigmaD.

\medskip
\noindent\textbf{Performance of LMT.}
We now evaluate the performance of the LMT algorithm in comparison to the optimal resolution 
in terms of computation time and total deviation. We compute the ratio $r=D(LMT,P)/D(S^*,P)$ where $S^*$ is a schedule returned by \sigmaD and $LMT$ is a schedule returned by the LMT algorithm. We compute $r$ before and after the local search. 

The LMT algorithm alone returns solutions with a sum of deviations about 6\% higher than the optimal sum of deviations. With local search, the solution improves and gets very close to the optimal solution, with on average a sum of deviation less than 1\% higher than the optimal one. In terms of computation time, for $10$ tasks and $100$ voters, the heuristic (LMT+local search) takes 0.037 seconds to return its solution before the local search, and 0.63 seconds in total, while the linear program takes 4.5 seconds. This heuristic is thus a very fast and efficient alternative at rule \sigmaD for large instances. 

\medskip
\noindent \textbf{Difference between the three rules.}
We execute the three rules on 300 instances, and we compare the schedules obtained with respect to  the total deviation ($\Sigma D$), the total tardiness ($\Sigma T$) and the weighted Kendall Tau score (KT). We compare each schedule obtained to the optimal schedule for the considered metric. For example, the ``1.06'' in column \sigmaT in  Table~\ref{tab:comparaison_r} means that, on average, for uniform instances with 5 tasks, the schedule returned by the \sigmaT rule has a sum of deviation 1.06 times larger than the minimum sum of deviation.

\begin{table}[H]
    \centering
    \begin{tabular}{c|c||c|c||c|c||c|c}
        \multicolumn{2}{c||}{} & \multicolumn{2}{c||}{\sigmaD} & \multicolumn{2}{c||}{\sigmaT} &  \multicolumn{2}{c}{PTA K}\\
        $P$ & $M$ & $n\!=\!5$ & $n\!=\!10$ & $n\!=\!5$  & $n\!=\!10$ & $n\!=\!5$ & $n\!=\!10$\\
        \hline
        \multirow{3}{*}{U} & \sigmaD & 1 & 1 & 1.06 & 1.07 & 1.07 & 1.09\\
         & \sigmaT & 1.12 & 1.16 & 1 & 1 & 1.01 & 1.02\\
         & KT & 1.12 & 1.16 & 1.01 & 1.01 & 1 & 1\\
        \hline
        \multirow{3}{*}{C} & \sigmaD & 1 & 1 & 1.05 & 1.09 & 1.05 & 1.07 \\
         & \sigmaT & 1.06 & 1.08 & 1 & 1 & 1.001 & 1.001\\
         & KT & 1.07 & 1.07 & 1.002 & 1.01 & 1 & 1\\
        \hline
    \end{tabular}
    \caption{Performance of each rule relative to the others.}
    \label{tab:comparaison_r}
\end{table}

Table~\ref{tab:comparaison_r} shows that the schedules returned by \sigmaT and PTA Kemeny are very close to each other (the values they obtain are very close),  
while the \sigmaD rule returns more different schedules, even if the scores obtained by the three rules do not differ from more than 16\% for uniform instances and 9\% for correlated instances.  
Note that the number of tasks does not seem to change these results. Overall,  the PTA Kemeny and \sigmaT rules return similar schedules, in which short tasks are favored, whereas the \sigmaD rule seems to return schedules as close as possible to the preference profile. 

\medskip
\noindent\textbf{Length reduction monotonicity (axiom LRM).}
We study to what extent the length reduction monotonicity axiom is fulfilled in practice. We run the three rules on 1200 instances with $50$ voters and $8$ tasks. Then, we reduce the length of a random task in each of the instances, and run the three rules again. If the reduced task starts later in the schedule returned by a rule than it did before the reduction, we count one instance for which the rule violates LRM. On the 1200 instances, PTA Kemeny and \sigmaT always respected LRM. The \sigmaD rule violated LRM in 102 instances (8.5\%). This percentage goes up to 12.3\% on uniform instances and up to 18\% on uniform instances with tasks with similar lengths.  

\section{Discussion and conclusion}
\label{sec:conclu}

In this paper, we showed that some standard axioms in social choice are not adapted to the collective schedule problem, and we introduced new axioms for tasks which have duration. These axioms may also be useful in some other contexts where candidates have weights.  We showed incompatibilities between axioms, showing that neutral or distance based rules are not PTA Condorcet consistent and do not approximate the sum of tardiness of the tasks. 

We also studied three aggregation rules for collective schedules, from an axiomatic and an experimental viewpoint. We saw that the PTA Kemeny and the \sigmaT rules seem to be particularly adapted in EB settings, whereas the \sigmaD rule is  useful in non EB settings.  
We conjecture that the PTA Kemeny and \sigmaT rules fulfill the length reduction monotonicity  axiom -- this is the case in our experiments but showing this from an axiomatic viewpoint is an open problem.


\medskip
\noindent \textbf{Acknowledgements}. We acknowledge a financial support from the project THEMIS ANR-20-CE23-0018 of the French National Research Agency (ANR).

%
%
%
\bibliographystyle{splncs04}
\bibliography{Bibliographie}

\end{document}